\documentclass{aa}  

\usepackage{graphicx,amsmath}
\usepackage{amssymb}
\usepackage{color}
\usepackage{natbib}		
\usepackage{url}
\usepackage{multirow}
\usepackage{threeparttable}      
\usepackage{soul}
\usepackage[varg]{txfonts}
\usepackage{xcolor}
\usepackage[normalem]{ulem}
\usepackage[colorlinks=true]{hyperref}

\graphicspath{ {./Graphics/} }

\bibpunct{(}{)}{;}{a}{}{,}           

\def\approxinf{%
  \def\p{%
    \setbox0=\vbox{\hbox{$<$}}%
    \ht0=0.6ex \box0 }%
  \def\s{%
    \vbox{\hbox{$\sim$}}%
  }%
  \mathrel{\raisebox{0.7ex}{%
      \mbox{$\underset{\s}{\p}$}%
    }}%
}

\newcommand{\bc}[1]{\textbf{#1}}

\begin{document}

\title{The JADE code: Coupling secular exoplanetary dynamics and photo-evaporation}

\author{
   O. Attia\inst{1} \and
   V. Bourrier\inst{1} \and
   P. Eggenberger\inst{1} \and
   C. Mordasini\inst{2} \and
   H. Beust\inst{3} \and
   D. Ehrenreich\inst{1}
	  }
   
\authorrunning{O.~Attia et al.}
\titlerunning{The JADE code.}

\institute{
Observatoire Astronomique de l'Universit\'e de Gen\`eve, Chemin Pegasi 51b, 1290 Versoix, Switzerland \\
e-mail: \url{omar.attia@unige.ch} \and
Department of Space Research \& Planetary Sciences, University of Bern, Gesellschaftsstrasse 6, 3012 Bern, Switzerland \and
Universit\'e Grenoble-Alpes, CNRS, IPAG, 38000 Grenoble, France
}
   
\date{Received 16 September 2020 / Accepted 23 December 2020}
 
\abstract{
Close-in planets evolve under extreme conditions, which raises questions about their origins and current nature. Two evolutionary mechanisms thought to play a predominant role are orbital migration, which brings them close to their star, and atmospheric escape under the resulting increased irradiation. Yet their relative roles remain poorly understood, in part because we lack numerical models that couple the two mechanisms with high precision and on secular timescales. To address this need, we developed the Joining Atmosphere and Dynamics for Exoplanets (JADE) code, which simulates the secular atmospheric and dynamical evolution of a specific planet around its star, and can include the perturbation induced by a distant third body. On the dynamical side, the three dimensional evolution of the orbit is modeled under stellar and planetary tidal forces, a relativistic correction, and the action of the distant perturber. On the atmospheric side, the vertical structure of the atmosphere is integrated over time based on its thermodynamical properties, inner heating, and the evolving stellar irradiation, which results, in particular, in extreme ultraviolet (XUV)-induced photo-evaporation.

The JADE code is benchmarked on GJ436 b, which is a prototype of the evaporating giants on eccentric, misaligned orbits at the edge of the hot Neptunes desert. We confirm previous results that the orbital architecture of GJ436 b is well explained by Kozai migration and bring to light a strong interplay between its atmospheric and orbital evolution. During the resonance phase, the atmosphere pulsates in tune with the Kozai cycles, which leads to stronger tides and an earlier migration. This triggers a strong atmospheric evaporation several billion years after the planet formed, refining the paradigm that mass loss is dominant in the early age of close-in planets. These results suggest that the edge of the desert could be formed of warm Neptunes whose evaporation was delayed by Kozai migration. They strengthen the importance of coupling atmospheric and dynamical evolution over secular timescales, which the JADE code will allow for one to simulate for a wide range of systems.
}


\maketitle

\hypersetup{citecolor=blue}
\hypersetup{linkcolor=blue}


\section{Introduction}

As of December 2020, more than 4300 exoplanets have been found, nearly half of which orbit in less than 10 days around their host star\footnote{\url{http://exoplanet.eu/}; \url{https://exoplanetarchive.ipac.caltech.edu/}}. This population of close-in planets ranges from small rocky objects to Jupiter-sized giants, even so it displays a surprising deficit of Neptune-sized planets on very short orbits \citep[$\approxinf$3\,days;][]{Lecavelier2007,Davis2009,Szabo2011,Beauge2013,Lundkvist2016,Mazeh2016}. This so-called desert of hot Neptunes is a key feature in exoplanet science, as the imprint of processes that determined the evolution and present nature of close-in planets. While the desert was the focus of many studies over the last decade, its origin and the properties of planets defining its borders remain unclear \citep[e.g.,][]{Mazeh2016,Zahnle2017,Owen2018}.

Hot Neptunes may have lost their atmosphere through evaporation, which is an efficient hydrodynamic escape driven by the stellar X-ray and extreme ultraviolet (XUV) radiation \citep[e.g.,][]{VidalMadjar2003,Lammer2003,Lecavelier2007,MurrayClay2009,Owen2012,Tripathi2015}. However, our understanding of atmospheric escape in irradiation conditions unmet in the Solar System remains limited by the lack of direct observations. For years, only hot Jupiters were observed losing their atmosphere \citep[e.g.,][]{VidalMadjar2003,Lecavelier2012}, at tremendous rates of thousands of tons per second that nonetheless do not affect the evolution of these massive planets \citep{Hubbard2007,Ehrenreich2011b}. The erosion of hot Neptunes, on the contrary, is thought to have formed a fraction of the mini-Neptunes and hot rocky planets at the lower-radius border of the desert \citep{Lecavelier2004,Owen2013,Lopez2013}. Alternatively, it was proposed that orbital migration played a role in shaping the desert, with different classes of planets forming differently \citep[e.g.,][]{Batygin2016,Mazeh2016,Lee2016} or following a different dynamical evolution \citep{Matsakos2016,Owen2018}. The dynamical history of a planet can be traced back in time by its present-day orbital architecture \citep[the shape and orientation of its orbit around its star, see review by][]{Triaud2018}, but most measurements have been obtained for Jupiter-sized planets around early-type stars, limiting our ability to assess the impact of migration on a large variety of systems. Understanding the origins of the different classes of planets around the desert and their possible filiation thus requires that we obtain observations of escape and architecture for a wider range of systems, and that we use them to inform models simulating their coupled atmospheric and dynamical evolution.

Indeed, when and how a planet migrates determines how much and how long it gets irradiated. Close-in planets that migrated within the proto-planetary disk are expected to lose their atmosphere early-on \citep[e.g,][]{Jackson2012,Jin2014}, as they migrate through the protoplanetary disk and get close to the young energetic host star \citep[e.g.,][]{Baruteau2013}. Such processes are further expected to maintain disk and planetary orbits aligned \citep[e.g.,][]{Marzari2009}. Yet some migration pathways can act over much longer time scales, as for secular migration \citep{Wu2011}, scattering \citep{Ford2008,Nagasawa2008,Nagasawa2011}, and Kozai migration \citep{Wu2003,Fabrycky2007,Naoz2011,Teyssandier2013b}, raising questions about the atmospheric evolution of a planet getting close to its star billions of years after its formation. These mechanisms can misalign planetary systems, offering the possibility to disentangle short- and long-term dynamical evolution. Many hot Jupiters, for example, are on misaligned orbits that would naturally result from high-eccentricity migration \citep{Naoz2012,Albrecht2012,Davies2014,Petrovich2016,Teyssandier2019}.

Recent discoveries brought new insights into these questions. Transit observations with the \textit{Hubble} Space Telescope led to the detection of giant clouds of hydrogen around two warm Neptunes located at the border of the desert as for GJ436 b \citep{Kulow2014,Ehrenreich2015,Lavie2017,DosSantos2019} and GJ3470 b \citep{Bourrier2018c}, supporting the atmospheric erosion of hot Neptunes as a key mechanism. GJ3470 b could already have lost up to half of its mass over its 2 Gyr lifetime and might keep eroding until it gives birth to a mini-Neptune \citep{Bourrier2018c}. GJ436 b is subjected to a smaller present-day mass loss than GJ3470 b \citep{Bourrier2015c,Bourrier2016}, but is surprisingly more massive (23 vs 13 M$_\mathrm{Earth}$) for its age (4 - 8 Gyr). Indeed, GJ436 b would likely have lost as much mass as GJ3470 b in its youth if it migrated early-on, raising questions about how and when it reached its present location. Interestingly, the orbit of GJ436 b is eccentric \citep[suggesting the planet has not been orbiting for long close to its M dwarf host,][]{Tong2009,Beust2012} and it was recently shown to be nearly polar \citep{Bourrier2018a}. Both features are naturally explained if the planet was brought close to its star after several billion years by gravitational interactions with an outer companion \citep{Bourrier2018a}. This late Kozai migration could have triggered the evaporation of GJ436 b, which would further explain its moderate erosion.

This led \citet{Bourrier2018a} to propose that GJ436 b is the prototype for a class of warm Neptunes that underwent orbital migration long after their formation and reached the fringes of the desert in recent times \citep{Owen2018}. These planets would either have already reached a stable orbit far enough from their star to be safe from evaporation, like GJ436 b, or could still be migrating and will erode as they move farther into the desert. The nonzero eccentricity of most warm Neptunes at the edge of the desert could be a consequence of this scenario \citep{Correia2020}, but an in-depth characterization of the orbital architecture and evaporation status of close-in planets all around the desert is needed to investigate this theory. This effort needs to be complemented by numerical models able to simulate the detailed secular evolution of close-in planets (i.e., on time scales comparable to the age of the system) under the coupled effects of atmospheric escape and orbital migration. The purpose of this paper is to present the JADE code, which was developed to address this need. In Sect. \ref{sect:desc}, we describe the structure of the model we have developed. In particular, we give a general overview of the JADE code, then we focus successively on the orbital and atmospheric features of the model. In Sect. \ref{sect:gj436}, we apply the JADE code to the particular case of GJ436 b to compare our results on its dynamical evolution with published studies and to investigate its coupling with atmospheric evolution. Finally, conclusions and perspectives are presented in Sect. \ref{sect:ccl}.


\section{Description of the JADE code}
\label{sect:desc}


\subsection{General description}

Various models have been designed to simulate the precise dynamical behavior of close-in planet systems, including long-term processes such as tidal friction or Kozai-Lidov oscillations \citep[e.g.,][]{Eggleton1998,Eggleton2001,Mardling2002,Fabrycky2007,Beust2012}. These models generally consider planets as particles or rigid spheres and thus do not take into account the planetary structure and its evolution (e.g., due to atmospheric mass loss or the cooling of the core).

On the other hand, atmospheric models have been developed for a wide range of exoplanets and purposes. They can be roughly separated between lower-atmosphere models, describing the layers up to the base of the thermosphere \citep[e.g.,][]{Guillot2002,Madhusudhan2009, Guillot2010,Heng2012,Komacek2017}, and models that simulate the expansion of the thermosphere and exosphere under the irradiation of the star \citep[e.g.,][]{Yelle2004,Owen2012,Bourrier2013b,Koskinen2013a,Salz2016a,Johnstone2018}. These models provide a detailed depiction of the atmosphere for a given state of the planetary system, but do not follow its evolution over long time scales. Models that were developed for this purpose, in particular to study the impact of evaporation, generally do not account for the evolution of the planetary orbit \citep[e.g.,][]{Lopez2012,Lopez2013,Owen2017,Lopez2017,Kubyshkina2018b}.

Studies that investigated the coupling between atmospheric and orbital evolution generally used simple approximations of dynamical processes to allow simulating the exoplanet population as a whole \citep[e.g.,][]{Kurokawa2014,Owen2018} or only accounted for migration in the early stages of the planetary system when evaporation is considered dominant \citep[e.g.,][]{Jin2014,Jin2018}. \citet{Barnes2020} recently developed a multi-purpose model that accounts for numerous atmospheric (e.g., mass loss, climate of Earth-like planets) and dynamical (e.g., gravitation in multiple-planet systems, tides, and galactic evolution) features. Yet their code lacks a layer-by-layer thermodynamical structure integrator and cannot simulate highly eccentric orbits, which prevents studying high-eccentricity migration pathways, in particular the Kozai-Lidov resonance.

\begin{figure*}
\centering
\includegraphics[width=17cm]{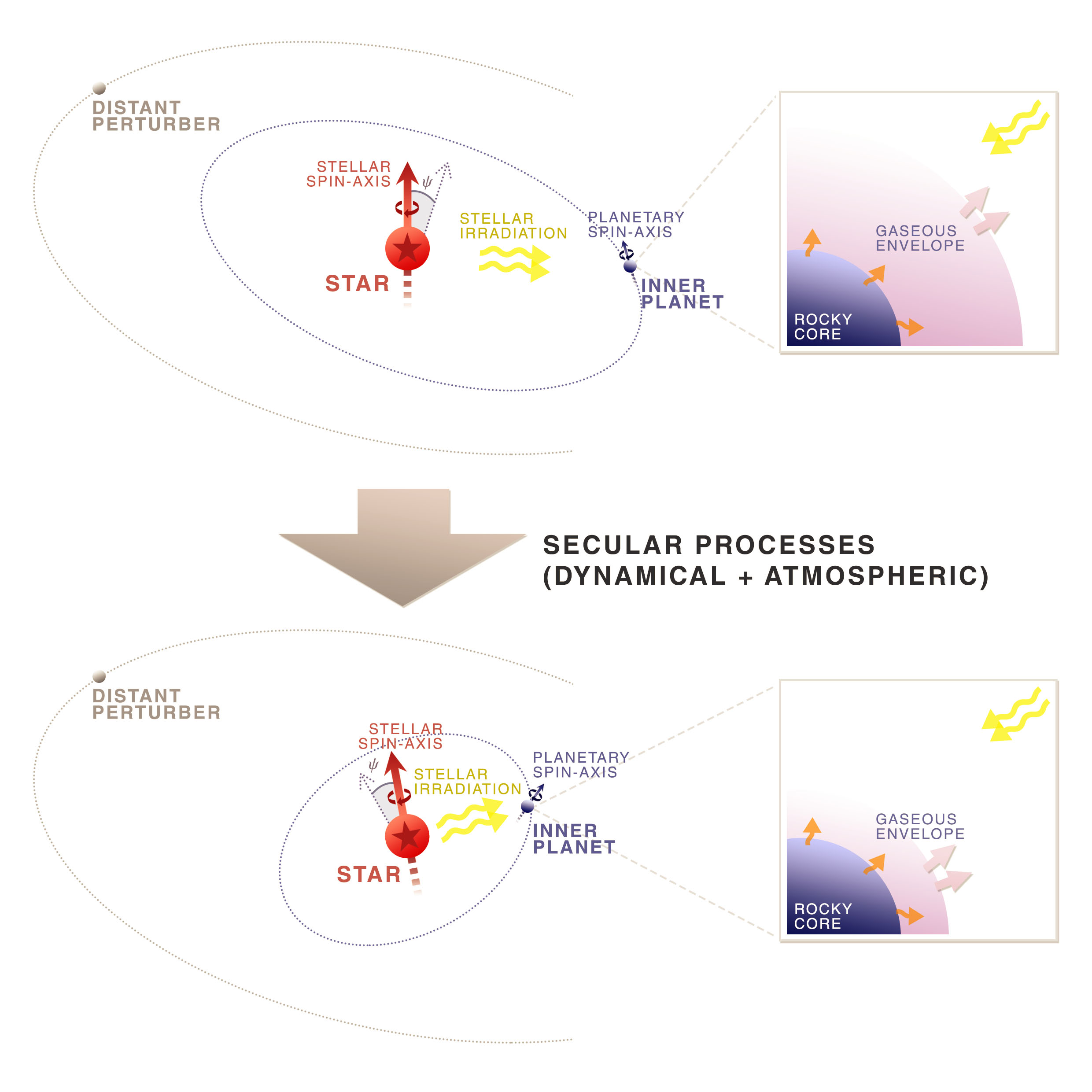}
\caption{Illustration of how the JADE code works. The central star and its spin-axis are in red. The stellar irradiation is represented in yellow. The inner planet orbiting the star is in blue, as well as its spin-axis. Its orbit is represented in dotted blue lines. The dotted blue arrow is the normal to the inner orbital plane. The spin-orbit angle $\psi$ is highlighted in light blue. The distant perturber orbiting the star is represented in gray, as well as its orbit in dotted lines. A zoom on the atmospheric structure of the inner planet is depicted in the right part of the figure. It is composed of a gaseous envelope atop a rocky core. The orange arrows represent the radiogenic heating from the core and the light-purple ones account for atmospheric evaporation. From top to bottom, the figure depicts the configuration of the system at two different secular time steps. Due to secular dynamical processes, the shape of the inner orbit, as well as the stellar and planetary spin-axes, vary over time. The figure illustrates a typical case where the inner orbit shrank, circularized, and changed inclination. Mass loss of the inner planet's gaseous envelope under the influence of stellar irradiation can lead to substantial changes in the planet's mass and atmospheric structure. On the other hand, the outer orbit is considered immutable in the present framework.}
\label{fig:jade}
\end{figure*}

With this necessity in mind, the JADE code combines an accurate secular dynamical integrator, a fully coherent atmospheric structure integrator, and XUV-driven photo-evaporation (Fig. \ref{fig:jade}). To best capture the interplay between atmospheric and orbital evolution, we devised a model coupling these processes from the bottom up, rather than joining independent modules based on existing dynamical and atmospheric models. We explicitly clarify all the equations governing the JADE code, so that its features are fully understandable and its results reproducible. The code simulates the evolution of a planet around its host star, with the possible addition of a second object on a distant orbit. We assume that this outer companion, which can represent a planet, a brown dwarf, or a star, does not feel the action of the inner planet and of the star so that it only acts as a perturber. The atmospheric evolution of the inner planet is influenced by stellar irradiation, which leads to mass loss via photo-evaporation, and by the cooling of the planet’s core. The JADE code monitors the secular evolution of the inner planet’s three dimensional (3D) orbital coordinates (in particular the angle between the stellar spin axis and the normal to the orbital plane, which is an important observational marker of migration), as well as its main atmospheric properties (such as the mass and radius of the gaseous envelope).

\subsection{Orbital features}

\begin{figure}
\resizebox{\hsize}{!}{\includegraphics{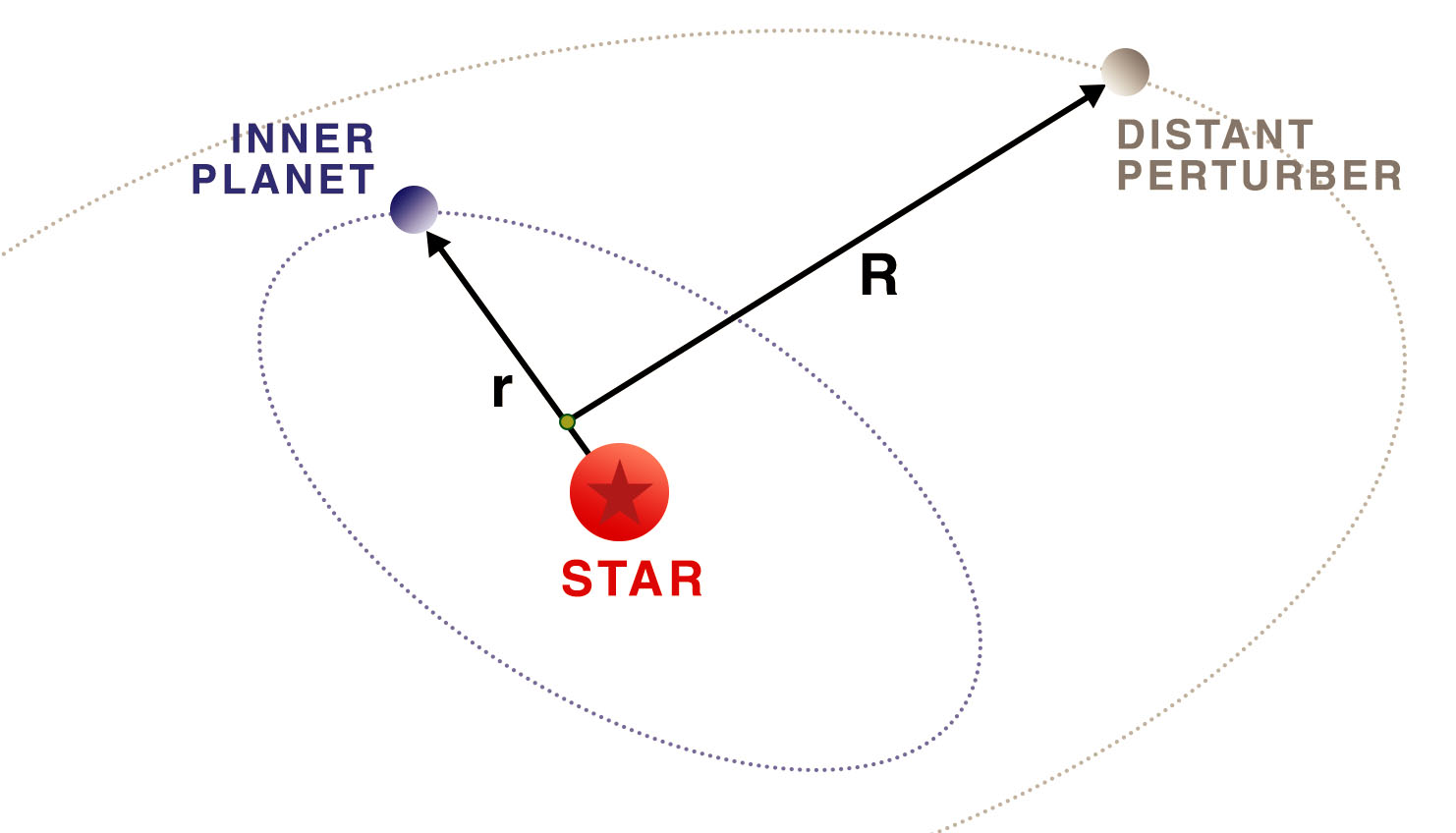}}
\caption{Illustration of the orbital configuration modeled by the JADE code. A main planet orbits around a central star. There is also the possibility of adding a distant perturber orbiting around the same star. The orbits are represented by the dotted lines. The vector $\mathbf{r}$ links the star to the inner planet, while $\mathbf{R}$ links the center of mass of the star and the inner planet to the distant perturber. They illustrate the hierarchical Jacobi coordinates used in this kind of dynamical problem.}
\label{fig:orbit}
\end{figure}

The dynamical mechanisms encoded in the JADE code are tidal forces, the influence of the distant perturber \citep[which can lead to drastic changes over secular time scales,][]{Mardling2002,Beust2012}, and a post-Newtonian relativistic correction. The inner and outer orbits are described with 3D coordinates and can thus be initialized to classical coplanar orbits as well as highly eccentric or mutually inclined orbits (a particularly important configuration for the Kozai resonance). Since the outer companion is considered as a perturber, its orbit is fixed and only the evolution of the inner orbit is calculated. We note that the perturber can even be removed entirely from the simulation, so that the code calculates the evolution of a single planet around its star. Figure \ref{fig:orbit} illustrates a generic orbital configuration.

\subsubsection{Orbital equations of motion}

Hierarchical (Jacobi) coordinates are the preferred coordinating system in our case. The principle of Jacobi coordinates is that the orbit of the innermost planet is referred to the star and the orbit of the second planet is referred to the center of mass of the star and the innermost planet. The other possibility would have been to adopt stellocentric coordinates, where all orbits are referred to the star. However, the Jacobi system has the advantage that the relative orbits are simply perturbed Keplerian orbits so that the orbital elements are easy to calculate \citep{Murray1999}. It was also shown by \citet{Beust2003} that once it is expressed using Jacobi coordinates, a hierarchical N-body system naturally splits into a collection of independent Keplerian orbits with perturbations that only depend on the planet positions, contrary to the use of stellocentric coordinates. This ensures a better stability of the numerical integrations. Figure \ref{fig:orbit} shows such coordinates for our model. The referential in which all calculations are made is the fixed rest frame of the observer: two unit vectors in the sky plane $\mathbf{\hat{i}}$ and $\mathbf{\hat{j}}$ and a vector along the line-of-sight $\mathbf{\hat{k}}$. The three vectors in this order form a direct orthonormal basis. We globally adopt the notation where $\mathbf{\hat{v}}$ is the unit vector in the direction of $\mathbf{v}$.

In this context, the equation governing the relative motion of the innermost pair is \citep{Mardling2002}:
\begin{equation}
\label{eq:rdd}
\ddot{\mathbf{r}}=-\frac{G(M_{\rm{s}}+M_{\rm{pl}})}{r^3}\mathbf{r}+\sum_{i}\mathbf{f_{i}}.
\end{equation}
Here, \textbf{r} is the vector linking the center of mass of the star to the center of mass of the innermost planet (and $r$ its norm), $G$ the gravitational constant, $M_{\rm{s}}$ the mass of the star, $M_{\rm{pl}}$ the mass of the inner planet. The accelerations $\mathbf{f_i}$ account for the perturbations to the Keplerian orbit. They are due to the perturber $\mathbf{f_{\rm pert}}$, the tidal and spin distorsions of the star (resp. the innermost planet) $\mathbf{f_{\rm SD}^{\rm s}}$ (resp. $\mathbf{f_{\rm SD}^{\rm pl}}$), the tidal damping of the star (resp. the innermost planet) $\mathbf{f_{\rm TD}^{\rm s}}$ (resp. $\mathbf{f_{\rm TD}^{\rm pl}}$), and the relativistic potential of the binary formed by the inner planet and the star $\mathbf{f_{\rm rel}}$. We recall their formulation in Appendix \ref{app:fi}. 

The spin rate of the star and the inner planet around themselves is subject to change due to the tidal forces. Assuming solid-body rotations, the evolution of the spin vectors are given by \citep{Mardling2002,Beust2012}:
\begin{equation}
\label{eq:Osd}
I_{\rm{s}}\dot{\mathbf{\Omega}}_{\mathbf{s}}=-\frac{M_{\rm{s}}M_{\rm{pl}}}{M_{\rm{s}}+M_{\rm{pl}}}\mathbf{r}\times(\mathbf{f_{SD}^{s}}+\mathbf{f_{TD}^{s}})
\end{equation}
\begin{equation}
\label{eq:Opd}
I_{\rm{pl}}\dot{\mathbf{\Omega}}_{\mathbf{pl}}=-\frac{M_{\rm{s}}M_{\rm{pl}}}{M_{\rm{s}}+M_{\rm{pl}}}\mathbf{r}\times(\mathbf{f_{SD}^{pl}}+\mathbf{f_{TD}^{pl}}),
\end{equation}
where $I_{\rm{s}}$ and $I_{\rm{pl}}$ are respectively the moment of inertia of the star and the main planet.

It is possible to directly integrate over time Eqs. (\ref{eq:rdd}), (\ref{eq:Osd}) and (\ref{eq:Opd}). This would fully characterize all the main planet's orbital elements by following the procedure outlined in \citet[p. 53]{Murray1999}. However, as these equations follow the evolution of the position, velocity and acceleration vectors, the integration step is necessarily controlled by the inner orbital period. Modeled planets can have periods in the range of only a few days. Hence, simulating secular time scales (billion years) using this approach and an integration step of a few days is inappropriate because of the computation time.

\subsubsection{Secularization}

A common solution for this kind of problems is ``secularization''. It relies on two main steps: a change of coordinate system and an averaging of the dynamical equations over the planetary orbits. The change of frame consists in no longer integrating the position $\mathbf{r}$ and velocity $\dot{\mathbf{r}}$ vectors. Instead, we follow the evolution of two slow-varying vectors: the orbital angular momentum per unit mass vector $\mathbf{h}$, a vector normal to the orbital plane, and the Runge-Lenz vector $\mathbf{e}$, a vector in the direction of periastron with magnitude equal to the orbital eccentricity \citep{Murray1999,Mardling2002,Beust2012}. The expressions of these vectors are given by:
\begin{equation}
\mathbf{h}=\mathbf{r}\times\dot{\mathbf{r}},\;\;\;\;\;\;\;\;\;\;
\mathbf{e}=\frac{\dot{\mathbf{r}}\times\mathbf{h}}{G(M_{\rm{s}}+M_{\rm{pl}})}-\mathbf{\hat{r}}.
\end{equation}

This procedure is like assimilating a planet to a torus of matter equally scattered on its orbit which is characterized by the two vectors $\mathbf{h}$ and $\mathbf{e}$. The new equations of motion, which are just the rate of change of $\mathbf{h}$ and $\mathbf{e}$, become:
\begin{equation}
\label{eq:hd}
\frac{\mathrm{d}\mathbf{h}}{\mathrm{d}t}=\sum_{i}\mathbf{r}\times\mathbf{f_i}
\end{equation}
\begin{equation}
\label{eq:ed}
\frac{\mathrm{d}\mathbf{e}}{\mathrm{d}t}=\frac{\sum_{i}[2(\mathbf{f_i}\cdot\dot{\mathbf{r}})\mathbf{r}-(\mathbf{r}\cdot\dot{\mathbf{r}})\mathbf{f_i}-(\mathbf{f_i}\cdot\mathbf{r})\dot{\mathbf{r}}]}{G(M_{\rm{s}}+M_{\rm{pl}})},
\end{equation} 
where the different $\mathbf{f_i}$ represent the forces listed above.

The set of differential equations that has to be integrated (Eqs. \ref{eq:Osd}, \ref{eq:Opd}, \ref{eq:hd} and \ref{eq:ed}) is then averaged over both orbital motions to remove the time dependency over their short periods. We note that this way we neglect any effect due to a potential mean-motion resonance between the two orbits. But the perturber is assumed to orbit the star sufficiently further away than the inner planet, so that the orbital period ratio is large enough to ensure that no strong resonance will play a significant role. We point out that for the sake of accuracy, we compute all the different force contributions without neglecting any of them. We do not assume a punctual planet or coplanar orbits as in \citet{Mardling2002} and we do not neglect the tidal forces acting on the star as in \citet{Beust2012}. The full details of the process of secularization are explicitly provided in Appendices \ref{app:avg} and \ref{app:sec} and a description of the numerical integrator can be found in Appendix \ref{app:numerical}.

Finally, the retrieval of the orbital parameters routinely-used in exoplanet analyses is all the easier using $\mathbf{h}$ and $\mathbf{e}$. In particular, the eccentricity $e$, semi-major axis $a$, inclination with respect to the sky-plane $i$, and 3D spin-orbit angle $\psi$ of the inner orbit are given by the following expressions:
\begin{equation}
e=\lVert\mathbf{e}\rVert
\end{equation}
\begin{equation}
\label{eq:a}
a=\frac{\lVert\mathbf{h}\rVert^2}{G(M_{\rm{s}}+M_{\rm{pl}})(1-e^2)}
\end{equation}
\begin{equation}
\cos{i}=\mathbf{\hat{k}\cdot\hat{h}}
\end{equation}
\begin{equation}
\cos{\psi}=\mathbf{\hat{\Omega}_s}\cdot\mathbf{\hat{h}}.
\end{equation}

The JADE code can thus monitor the evolution of \bc{$\psi$}, which is a crucial tracer of the dynamical evolution of a planetary system \citep[e.g.,][]{Bourrier2018a,Teyssandier2019}. We note that another strong added-value of the JADE code is accounting for the stellar spin variations generated by planetary tides acting on the star \citep[e.g.,][]{Lai2012,Rogers2013,Xue2014}. Together with variations in the orbital inclination, they determine entirely the true evolution of the spin-orbit angle.

The total angular momentum should be conserved. Its expression is \citep{Mardling2002}:
\begin{equation}
\label{eq:J}
\mathbf{J}=\frac{M_{\rm{s}}M_{\rm{pl}}}{M_{\rm{s}}+M_{\rm{pl}}}\mathbf{h}+\frac{(M_{\rm{s}}+M_{\rm{pl}})M_{\rm{pert}}}{M_{\rm{s}}+M_{\rm{pl}}+M_{\rm{pert}}}\mathbf{H}+I_{\rm{s}}\mathbf{\Omega_s}+I_{\rm{pl}}\mathbf{\Omega_{pl}}.
\end{equation}
$\mathbf{H}=\mathbf{R}\times\dot{\mathbf{R}}$ is the angular momentum of the outer orbit. Analytically, $\dot{\mathbf{J}}$ should be 0 but if the calculation is done with the above-mentioned contributions, one can see that there will be remaining non-null terms corresponding to $\mathbf{f_{pert}}$. These remaining terms should be cancelled by the contribution from $\dot{\mathbf{H}}$, but we assumed an invariant outer orbit. Nevertheless, the condition that $\mathbf{J}$ remains constant is still satisfied if $a_{\rm{pert}} \gg a$ and if the perturber's eccentricity $e_{\rm{pert}}$ is not too high (i.e., if on average the inner planet does not come too close from the perturber). In this case, which we assumed from the beginning, the outer orbit acts like an ``angular momentum reservoir''. In all cases, the JADE code monitors the total angular momentum to check whether it actually remains constant or not.

\subsubsection{Validation of the orbital model}

\begin{figure}
\resizebox{\hsize}{!}{\includegraphics{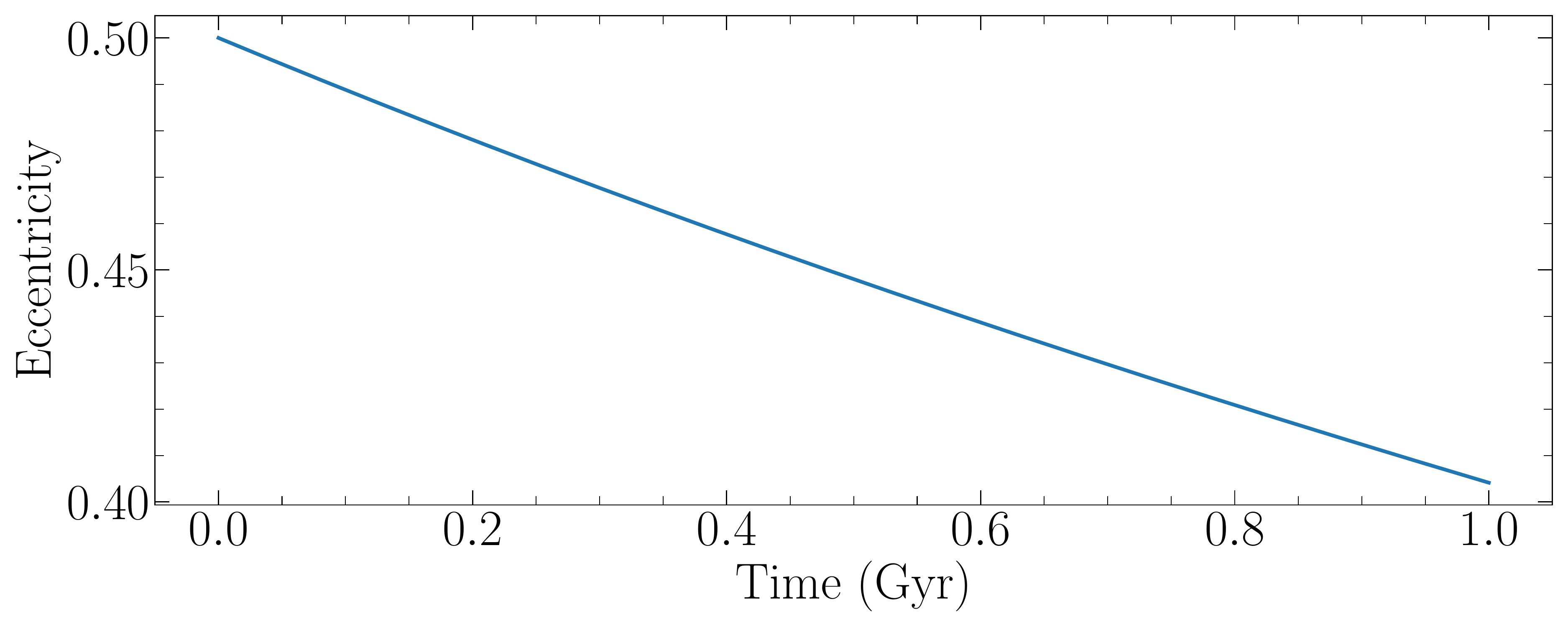}}
\resizebox{\hsize}{!}{\includegraphics{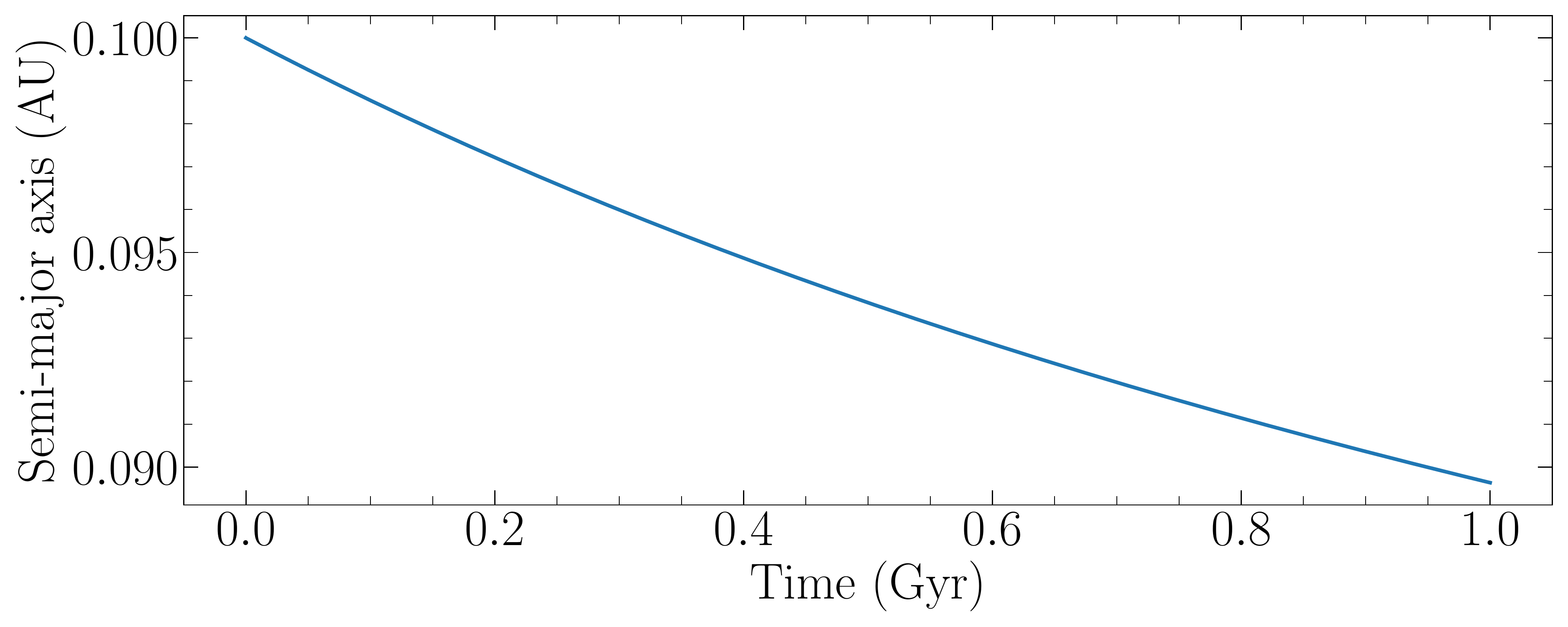}}
\caption{Secular evolution over 1 Gyr of the eccentricity (\textit{top}) and semi-major axis (\textit{bottom}) of the test system with non-null radii. The assumed parameters are: $M_{\rm s}=1M_\odot$,  $R_{\rm s}=1R_\odot$, $M_{\rm pl}=1M_{\rm Nept}$, and $R_{\rm pl}=1R_{\rm Nept}$. The initial values are: $e=0.5$, $a=0.1$ AU, $i=10^\circ$, and $\psi=0^\circ$.}
\label{fig:tides}
\end{figure}

The first obvious tests the JADE code had to pass were classical two-bodies tests without any relativistic correction. We started by successfully checking that the orbital parameters do not change regardless of the simulated system's age in case of a Keplerian orbit (i.e., a single punctual planet orbiting a punctual star). Afterwards, we allowed for non-null radii, which only triggers tidal effects. Following \citet{Beust2012}, Fig. \ref{fig:tides} shows the result of simulating a Neptune around a Sun at a distance of 0.1 AU for 1 Gyr. There is currently little knowledge about the tidal dissipation factor $Q_{\rm pl}$ that intervenes in the tidal contributions of the equations of motion. Even in the giant planets of the Solar System, $Q_{\rm pl}$ is only known within one order of magnitude at best. Hence, we assume a constant $Q_{\rm pl}$ for the inner planet during a simulation, even though it could potentially change substantially with the evolution of the planet's atmospheric structure and bulk properties. We use a value of $Q_{\rm pl}=10^4$ since the inferred value of the tidal dissipation factor for Neptune is in the range $9 \times 10^3 - 3.6 \times 10^4$ \citep{Zhang2008}. We assume a constant tidal factor $Q_{\rm s}$ for the star as well and use $Q_{\rm s}=10^5$ as a typical value, consistently with observational data regarding the circularization periods of binary stars \citep{Barker2009}. As expected, the eccentricity $e$ and the semi-major axis $a$ drop as a result of tidal damping and circularization. Furthermore, the spin-orbit angle, initialized so that the stellar spin is orthogonal to the orbital plane ($\mathbf{\Omega_{s}} \parallel \mathbf{\hat{h}}$), remains null. This is coherent with \citet{Lai2012} who concluded that the stellar spin-axis can only change direction due to tides generated by the planet on the star if the system is not initially aligned. While the stellar spin-axis can be brought to evolve because of the tides generated by the planet, we note that the JADE code does not include for now other processes leading to stellar rotational effects, such as shear instabilities or the magnetic braking of the stellar surface. Finally, the relative angular momentum error stays satisfactorily very low $< 10^{-14}$.

\begin{figure*}
\centering
\includegraphics[width=8.5cm]{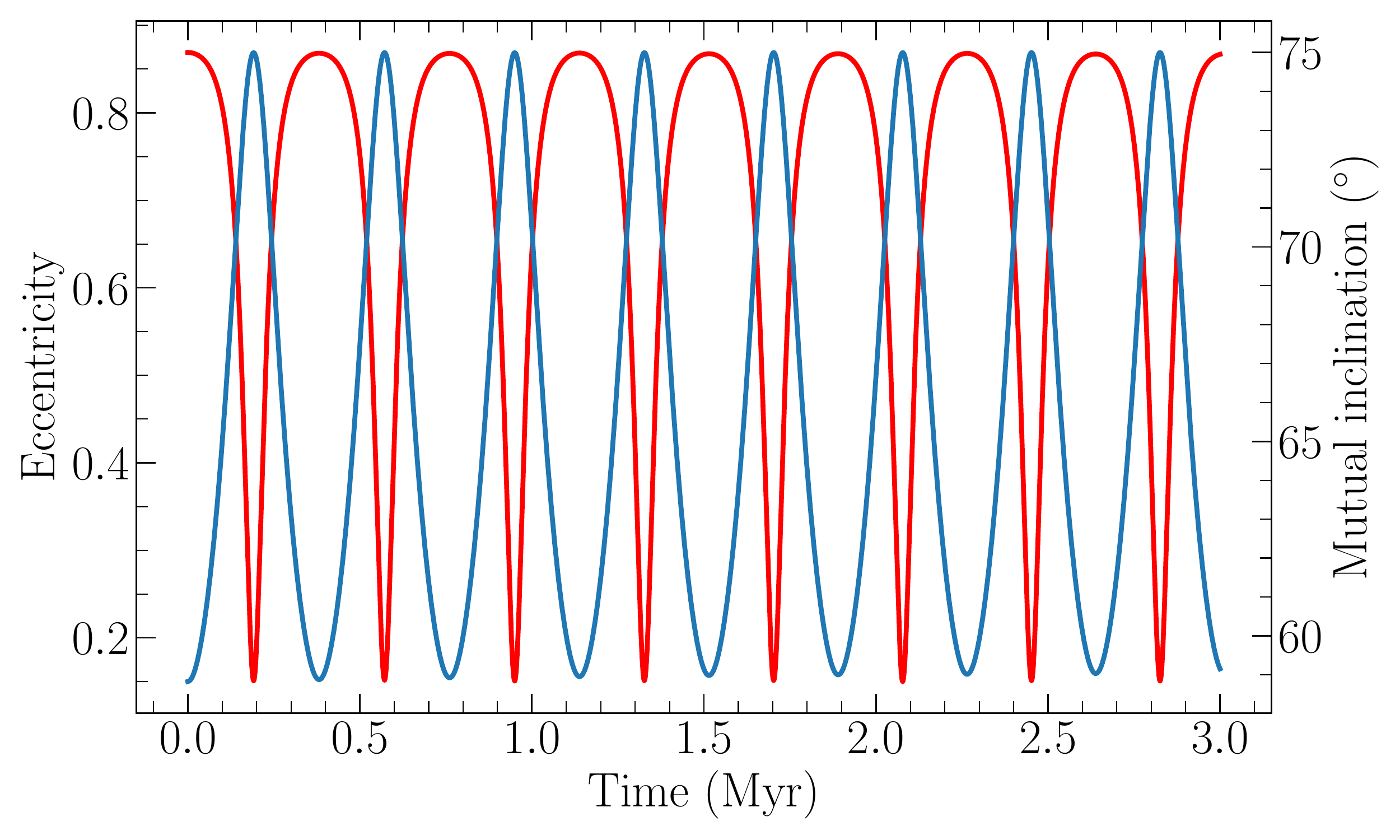}
\includegraphics[width=8.5cm]{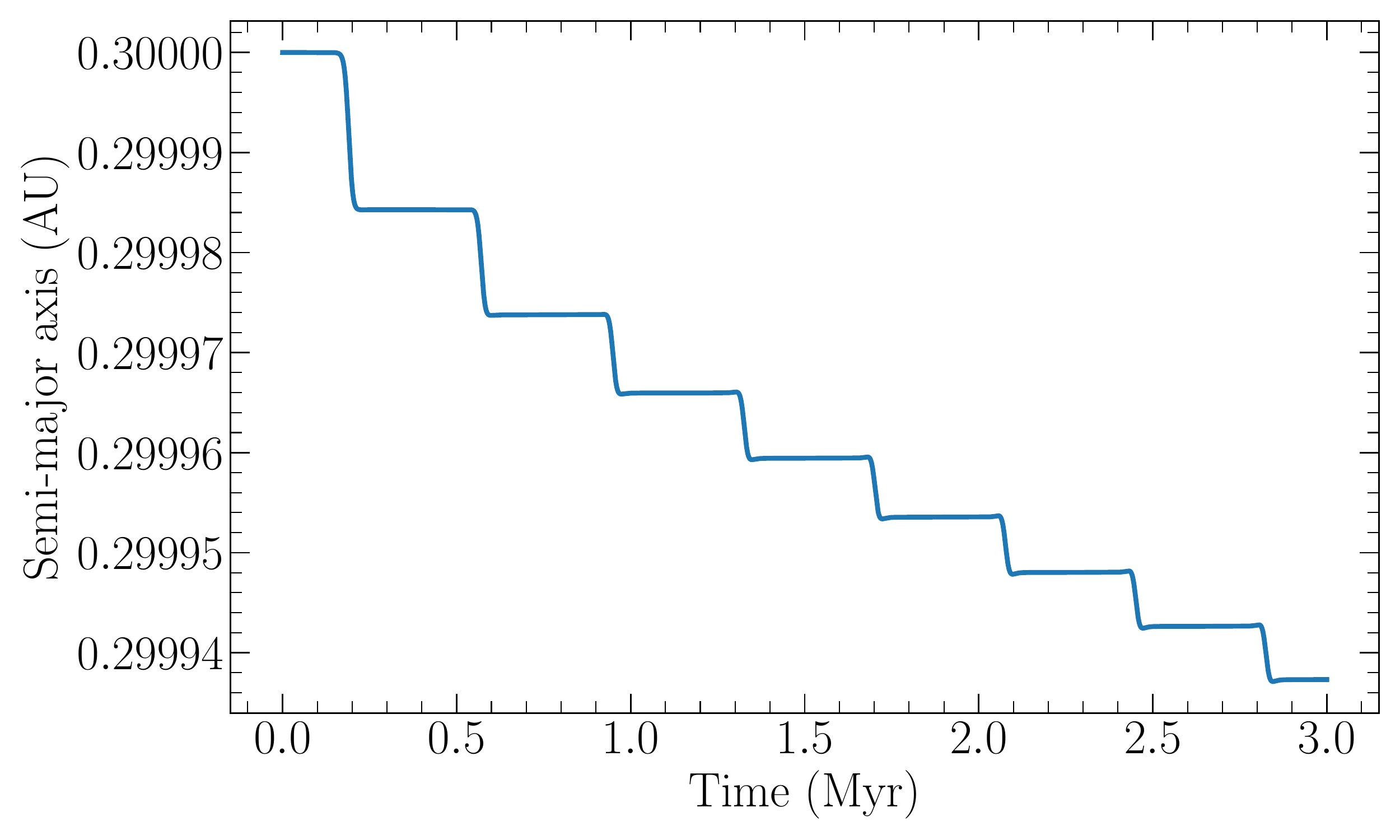}
\caption{Secular evolution over 3 Myr of the test system with non-null radii and a distant perturber. The assumed parameters are: $M_{\rm s}=1M_\odot$,  $R_{\rm s}=1R_\odot$, $M_{\rm pl}=0.06M_{\rm jup}$, $R_{\rm pl}=0.4R_{\rm jup}$, and $M_{\rm pert}=1M_{\rm jup}$. The initial values are: $e=0.15$, $a=0.3$ AU, $i=0^\circ$, and $\psi=0^\circ$ for the main planet and $e_{\rm pert}=0.01$, $a_{\rm pert}=5$ AU, and $i_{\rm pert}=75^\circ$ for the perturber. \textit{Left}: eccentricity (blue) and mutual inclination (red). \textit{Right}: semi-major axis.}
\label{fig:kozai}
\end{figure*}

The JADE code was specifically designed with the intention of simulating hierarchical three-body systems. The study of long-term stability in such systems has been thoroughly investigated \citep[e.g.,][]{Mazeh1979,Asanova1986,Eggleton1995,Kiseleva1998,Naoz2013}. In most cases, it requires that the perturber remains far enough and with a low enough eccentricity so as not to approach the inner orbit too much. However, even such weak perturbations can have important secular effects on the inner orbit when the inner and the outer orbits are not aligned \citep{Eggleton2001,Fabrycky2007,Beust2012,Bourrier2018a}. Under some circumstances, a perturber can generate Kozai-Lidov oscillations. It is a dynamical mechanism affecting the orbit of a binary system (star + inner planet) perturbed by a distant third body causing a periodic exchange between the inner orbit's eccentricity and inclination over secular time scales \citep{Kozai1962,Lidov1962}. In a three-body system, the orbit-averaged equations of motion for the inner planet have a conserved quantity thanks to axial symmetry: the component of the planet's orbital angular momentum parallel to the angular momentum of the star \citep{Merritt2013}. It can be expressed in terms of the eccentricity and the mutual inclination (difference between the inner planet's and the perturber's inclinations):

\begin{equation}
L_{\rm{}z}=\sqrt{1-e^2}\cos{}i_{\rm{}mut}.
\end{equation}

The mutual inclination is directly calculated in the JADE code as the angle between the inner and the outer orbits $\cos{i_\mathrm{mut}}=\mathbf{\hat{h}}\cdot\mathbf{\hat{H}}$. $L_{\rm z}$ being constant means that eccentricity can be traded for inclination: near-circular, highly inclined orbits can become very eccentric. The Kozai mechanism regime is only valid for small values of $L_{\rm z}$ \citep{Beust2012}, that is to say high initial mutual inclinations. It can be shown analytically \citep{Kinoshita1999} that the minimum inclination required for Kozai mechanism to start is the following:

\begin{equation}
\label{eq:condkozai}
i_{\rm{}mut}>\arccos{\sqrt{3/5}}\simeq39.2^\circ.
\end{equation}

However, this is true only for near-zero initial values of the eccentricity. Kozai mechanism can operate at smaller inclination if the initial eccentricity is larger; and at some point in the secular evolution, it will reach an inclination above this threshold.

In order to validate the JADE code regarding the Kozai mechanism, we simulate a $M_{\rm pl}=0.06M_{\rm jup}$, $R_{\rm pl}=0.4R_{\rm jup}$ Neptune around the Sun at $a=0.3$ AU and with an eccentricity $e=0.15$, perturbed by a distant Jupiter at $a_{\rm pert}=5$ AU with a low eccentricity $e_{\rm pert}=0.01$. Here, the lines of nodes of the two orbits are equal so that the mutual inclination is unambiguously related to the two individual inclinations $i_{\rm mut}=\lvert{}i-i_{\rm pert}\rvert{}$. In a first test, the inner and outer orbits are set aligned. As expected, this configuration fails to start the resonance: the eccentricity, semi-major axis and inclination remain constant over the simulated 3 Myr. We then set $i_{\rm mut}=75^\circ$ to check that Kozai cycles are triggered when the condition in Eq. (\ref{eq:condkozai}) is met. As can be seen in Fig. \ref{fig:kozai}, the eccentricity and inclination of the inner orbit indeed show the expected oscillations in the opposite direction to each other. So as to maintain $L_{\rm z}$ constant, increases in eccentricity (low values of $\sqrt{1-e^2}$) must be compensated by decreases in inclination (high values of $\cos{i_{\rm mut}}$). We notice that the resonance drives an initially weakly eccentric orbit to values of high eccentricity. Kozai cycles also induce a particular stair-shaped pattern in the variations of the semi-major axis. During the high eccentricity phases, tidal effects are stronger as the planet spends more time close to the star at the periastron, resulting in abrupt drops of the separation. Finally, the relative angular momentum error satisfyingly does not exceed $\simeq 10^{-6}$.

A final test is related to the characteristic time scale of Kozai oscillations \citep{Kiseleva1998,Merritt2013}:

\begin{equation}
\label{eq:taukozai}
\tau_{\rm{}Kozai}=\frac{2P_{\rm{}pert}^2}{3\pi{}P}\frac{M_{\rm s}+M_{\rm pl}+M_{\rm{}pert}}{M_{\rm{}pert}}(1-e_{\rm{}pert}^2)^{3/2},
\end{equation}

where $P_{\rm{}pert}$ and $P$ are respectively the orbital periods of the perturber and the inner planet, derived from Kepler's third law. $M_{\rm{}pert}=1M_{\rm jup}$ is the perturber's mass. Equation (\ref{eq:taukozai}) gives $\tau_{\rm Kozai}=1.68\times10^5$ yr for our test simulation, which is reasonably close to the value of the oscillation period considering that this formula is accurate within a factor on the order of unity \citep{Beust2006}.


\subsection{Atmospheric features}

The novelty of the JADE code compared to integrators that implement three-body exoplanetary dynamics is having the possibility of modeling an evolving planetary atmosphere. The inner planet's structure consists in a rocky core of mass $M_{\rm{}core}$ and a gaseous envelope of mass $M_{\rm{}env}$, so that the total planetary mass is $M_{\rm pl}=M_{\rm core}+M_{\rm env}$. The planetary radius $R_{\rm{}pl}$ is self-consistently integrated so as to be compatible with the planet's composition. The total mass and radius are the ones involved in the dynamical processes (i.e., used to solve the motion equations). Any change in the atmospheric structure of the simulated planet will thus have an impact on its dynamical evolution.

\subsubsection{Integration of the atmospheric structure}
\label{sect:atmostruc}

\begin{figure}
\resizebox{\hsize}{!}{\includegraphics{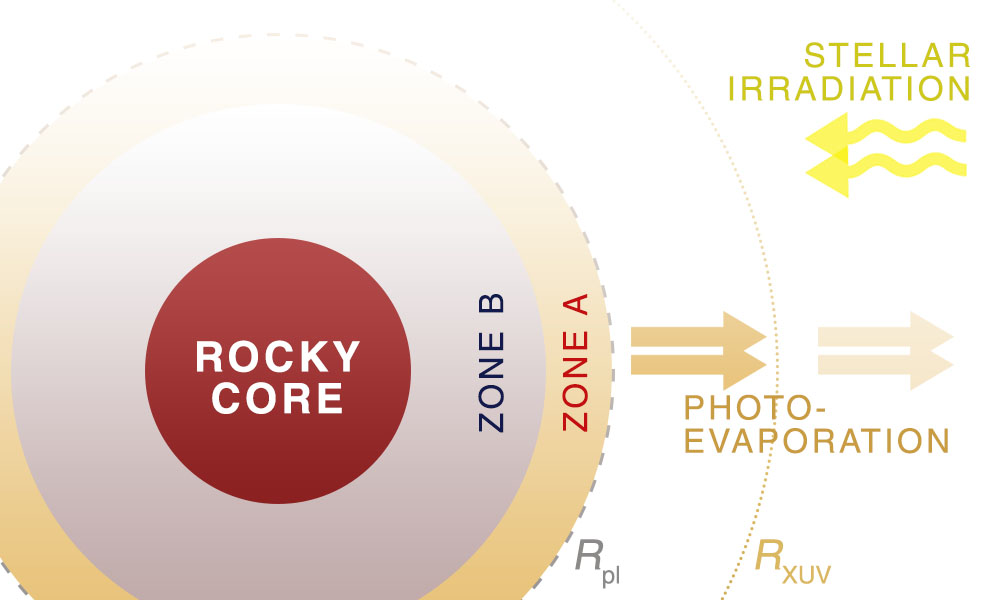}}
\caption{Illustration of the atmospheric structure of the main planet, as modeled by the JADE code. The planet corresponds to the rocky core plus the envelope, which consists in Zone A and Zone B. The XUV radius is above the optical radius and corresponds to the radius up to which the gas is lifted from Zone A so as to escape.}
\label{fig:atmostruct}
\end{figure}

The JADE code determines the radius of the optical photosphere $R_{\rm pl}$ with a similar approach as \citet{Lopez2013,Jin2014}. We consider a gaseous envelope dominated by hydrogen and helium, characterized by its time-dependent mass $M_{\rm env}(t)$ and constant helium mass fraction $Y_{\rm He}$, on top of a rocky core of a certain composition and with a certain constant mass $M_{\rm core}$. At well-chosen time steps of the simulation, the structure of the planet, set by the thermodynamical properties of the H/He mixture and the composition of the core, is fully integrated. This approach allows us to account for the evolution of the planet mass and radius with more accuracy than simple analytical relations \citep[e.g.,][]{Barnes2020}. We separate the envelope into the upper layers, where most of the optical and infrared stellar irradiation is absorbed (Zone A), and the lower layers, which are opaque to the stellar flux (Zone B). Including the rocky core, there are thus three separate regions that are integrated by the JADE code. Throughout the simulation, the atmosphere can evolve under photo-evaporative mass loss (driven by the stellar irradiation) and changes in the thermal structure (driven by stellar irradiation and the intrinsic planetary luminosity). Figure \ref{fig:atmostruct} illustrates the planetary structure.

We solve the atmospheric structure by assuming spherical symmetry and combining the one-dimensional hydrostatic equations \citep{Broeg2009,Mordasini2012,Jin2014}:

\begin{equation}
\label{eq:dmdr}
\frac{\mathrm{d}m}{\mathrm{d}r}=4\pi{}r^2\rho
\end{equation}
\begin{equation}
\label{eq:dpdr}
\frac{\mathrm{d}P}{\mathrm{d}r}=-\frac{Gm}{r^2}\rho
\end{equation}
\begin{equation}
\frac{\mathrm{d}\tau}{\mathrm{d}r}=-\kappa_{\rm{}th}\rho,
\end{equation}

where $r$ is the radius as measured from the planetary center, $m$ is the cumulative mass inside $r$, $\rho$ is the density in each elementary spherical shell, $P$ the pressure, $\tau$ the radially integrated optical depth from the center of the planet in thermal wavelengths and $\kappa_{\rm th}$ the thermal opacity. We interpolate $\kappa_{\rm th}$ from the \citet{Ferguson2005} Rosseland-mean opacity tables\footnote{``Caffau et al. 2011'' from \url{https://www.wichita.edu/academics/fairmount_college_of_liberal_arts_and_sciences/physics/Research/opacity.php}} at the helium fraction of the atmosphere, the temperature, and the density of each elementary layer. $\rho$ in the atmosphere is determined using the equation of state of \citet{Saumon1995} for a H/He gas for the temperature $T$ and the pressure $P$ of each elementary layer. See Appendix \ref{app:saumon} for more details.

The atmospheric temperature profile depends on both the optical depth and heat transfer mechanism (convective or radiative). In Zone A, we adopt the globally averaged temperature profile in the semi-gray model of \citet{Guillot2010}, derived using the Eddington approximation:

\begin{multline}
\label{eq:semigray}
T^4=\frac{3T_{\rm{int}}^4}{4}\left\{\frac{2}{3}+\tau\right\}+\frac{3T_{\rm{eq}}^4}{4}\Bigg\{\frac{2}{3}+\frac{2}{3\gamma}\left[1+\left(\frac{\gamma\tau}{2}-1\right)e^{-\gamma\tau}\right]+ \\
\frac{2\gamma}{3}\left(1-\frac{\tau^2}{2}\right)E_2(\gamma\tau)\Bigg\}.
\end{multline}

Here, $\gamma=\kappa_{\rm{v}}/\kappa_{\rm{th}}$ is the ratio of the visible opacity to the thermal opacity. $E_2(\gamma\tau)$ is the exponential integral $E_n(z)\equiv\int_1^\infty t^{-n}e^{-zt}\mathrm{d}t$. $T_{\rm int}$ and $ T_{\rm eq}$ are respectively the intrinsic temperature that characterizes the heat flux from the rocky core and the equilibrium temperature obtained from a balance with incident stellar energy. Their expressions are the following \citep{Chandrasekhar1939,Guillot2005,Jin2014}:

\begin{equation}
\label{eq:temp}
T_{\rm{int}}=\left(\frac{L_{\rm{}pl}}{4\pi\sigma_{\mathrm{B}}R_{\rm{}pl}^2}\right)^{1/4},\;\;\;\;\;\;\;\;\;\;
T_{\rm{eq}}=\left(\frac{L_{\rm{}bol}}{16\pi\sigma_{\rm{}B}\sqrt{1-e^2}a^2}\right)^{1/4},
\end{equation}

where $L_{\rm bol}$ represents the stellar bolometric luminosity and $\sigma_{\rm B}$ the Stefan-Boltzmann constant. The two temperatures are derived from a black-body law. The JADE code interpolates the value of $\gamma$ from the tabulated values of \citet{Jin2014} as a function of the equilibrium temperature. Note the $\sqrt{1-e^2}$ factor in the expression of $T_{\rm eq}$ resulting from the secularization of the bolometric flux (Appendix \ref{app:avg}).

In Zone B, the temperature gradient only depends on the heat transfer mechanism \citep{Broeg2009,Mordasini2012}:

\begin{equation}
\frac{\mathrm{d}T}{\mathrm{d}r}=\frac{T}{P}\frac{\mathrm{d}P}{\mathrm{d}r}\min\left(\nabla_{\rm{rad}},\nabla_{\rm{conv}}\right),
\end{equation}

where $\nabla_{\rm{rad}}$ is the radiative gradient and $\nabla_{\rm{conv}}$ is the convective gradient. Here, the Schwarzschild criterion \citep{Schwarzschild1958} is used to decide whether the energy transport occurs via radiative diffusion or convection. Any layer in which the gradient needed to transport the entire energy by radiation is larger than the convective gradient is considered convectively unstable (i.e., spontaneously generates convection cells) and the temperature gradient is set to $\nabla_{\rm conv}$. On the contrary, if $\nabla_{\rm rad} < \nabla_{\rm conv}$, the gas is convectively stable and we assume that the layer is entirely radiative. There is no convection and radiation mix within a same elementary layer (zero-entropy gradient hypothesis). In practice, the radiative gradient grows quickly with high pressures and $\nabla_{\rm rad}$ is chosen only for a thin layer in the upper part of Zone B. The radiative gradient is given by \citep{Mihalas2013,Jin2014}

\begin{equation}
\nabla_{\rm{rad}}=\frac{3\kappa_{\rm{th}}L_{\rm{}pl}P}{64\pi\sigma_{\rm{}B}GmT^4},
\end{equation}

while the convective gradient $\nabla_{\rm{conv}}$ is calculated using the equation of state of \citet{Saumon1995} for a H/He gas knowing $T$ and $P$. See Appendix \ref{app:saumon} for more details.

The integration of the rocky core is carryed out using Eqs. (\ref{eq:dmdr}) and (\ref{eq:dpdr}). The density profile is described by a polytropic equation of state that is temperature-independent. The core's density is given as a function of the pressure by \citep{Seager2007}

\begin{equation}
\rho(P)=\rho_0+cP^n,
\end{equation}

where the parameters $\rho_0$, $c$ and $n$ depend on the core's composition. This prescription used for the core is too simple to efficiently account for the detailed structure of rocky planets but is satisfactory for the interiors of Neptunes. A future version will extend the code to mini-Neptunes and smaller planets with a better description of the core and a greater variety of envelope compositions.

We assume that the luminosity of the planet is homogeneous \citep{Mordasini2012,Jin2014}. Although this assumption is not self-consistent with our layer-based model, it is valid at first order given that $L_{\rm pl}$ is concretely only used in a small region of the planetary atmosphere (Zone A and a thin upper layer in Zone B). Still, its value varies with time and is updated at each time step. We use the biquadratic fit in the core's and envelope's mass derived by \citet{Mordasini2020}.

\begin{multline}
\label{eq:Lpl}
\frac{L_{\rm{}pl}}{L_{\rm{jup}}}=a_0+b_1\left(\frac{M_{\rm{core}}}{M_{\rm{Earth}}}\right)+b_2\left(\frac{M_{\rm{core}}}{M_{\rm{Earth}}}\right)^2+c_1\left(\frac{M_{\rm{env}}}{M_{\rm{Earth}}}\right)+ \\
c_2\left(\frac{M_{\rm{env}}}{M_{\rm{Earth}}}\right)^2.
\end{multline}

$a_0$, $b_1$, $b_2$, $c_1$, and $c_2$ depend on time and are interpolated from the tabulated values of \citet{Mordasini2020}. The calculated $L_{\rm{pl}}$ includes the cooling and contraction of both the core and the envelope, as well as the radiogenic heat released in the interior of the planet due to the radioactive decay of the core. We use Eq. (\ref{eq:Lpl}) as a reasonable approximation because of the low dependence of $R_{\rm{pl}}$ on the planetary luminosity. Even though the $M_{\rm env}$ grid in \citet{Mordasini2020} does not extend to Jupiter-mass planets, they showed that Eq. (\ref{eq:Lpl}) yields acceptable values of $L_{\rm pl}$ for this class of planets. We will refine the calculation of $L_{\rm pl}$ in future versions of the JADE code to simulate more accurately a wider range of planets. The role of internal luminosity, in particular when accouting for additional sources such as ohmic dissipation \citep{Pu2017} or processes such as core-powered mass loss \citep{Ginzburg2018}, could be particularly critical for small planets and the formation of the radius valley \citep{Fulton2017}.

The integration of the planetary profiles is made from the top of the atmosphere to the center of the planet. We assume a certain planetary radius at the top boundary $R_{\rm pl}^{\rm hyp}$ and stop the integration when we reach the center ($r=0$). The JADE code runs a grid of models over a range of assumed radii $R_{\rm pl}^{\rm hyp}$ and retains the value of the radius that gives the closest calculated mass at the center $M_{\rm 0}^{\rm calc}$ to zero (see Appendix \ref{app:numerical} for more details).

For Zone A, integration starts with $m=M_{\rm pl}$, $P=1$ mbar, $\tau=2/3$, and $r=R_{\rm pl}^{\rm hyp}$. The initial value of $T$ is directly calculated using Eq. (\ref{eq:semigray}). The initial value of $\tau$ is the one usually taken to define a planetary radius when the planet is gaseous \citep[e.g.,][]{Jin2014,Lopez2017}. The initial value of $P$ is a standard value for the atmospheric pressure at such optical depth \citep{Jin2014} and we checked that the results of the simulations are largely insensitive to the initial value of $P$, independently of the envelope and core masses. The boundary between Zone A and Zone B should be at the optical depth in visible wavelengths. In other words, the optical depth should then satisfy $\tau\gg1/(\sqrt{3}\gamma)$ \citep{Rogers2011} based on the definition of $\gamma$ in Eq. (\ref{eq:semigray}). Hence, we set the limit between Zone A and Zone B at $\tau=100/(\sqrt{3}\gamma)$ \citep{Jin2014}.

\subsubsection{Photo-evaporation}
\label{sect:evap}

Close-in planets can lose a substantial fraction of their mass due to hydrodynamical escape induced by the high-energy input from the star \citep{VidalMadjar2003,Lammer2003,Lecavelier2007}. To simulate this process, one must determine the atmospheric mass-loss rate over the course of the planet life.

Various approaches have been used in the exoplanet literature to calculate mass-loss rates from atmospheric simulations, independently of observations \citep[as to the opposite case, see e.g.,][]{Bourrier2013b,Kislyakova2014}. One possibility consists in self-consistently simulating the full structure of the expanding upper atmosphere \citep[e.g.,][]{Salz2015,Salz2016b}, but it represents an impractical solution over long time scales. A complementary approach relies on pre-calculating a grid of upper atmosphere models to then interpolate the mass-loss rate at the desired planetary and stellar parameters \citep[e.g.,][]{Kubyshkina2018b}. However, this solution requires performing heavy preliminary simulations over a wide range of parameter space and is not as accurate as a direct calculation of the atmospheric structure. In the JADE code, we chose the more standard approach used in planetary evolution models, which consists in calculating mass-loss rates from an analytical formula based on the energy-limited escape \citep[e.g.,][]{Watson1981,Lammer2003,Lecavelier2004,Erkaev2007,Jin2014,Salz2016a,Lopez2017}. In this regime, the XUV photosphere (i.e., the layer where the planet becomes optically thin to ionizing photons) undergoes photo-evaporation due to strong X-rays and UV stellar irradiation and the mass-loss rate reads

\begin{equation}
\label{eq:massloss}
\dot{M}_{\mathrm{env}}=\epsilon\frac{L_{\rm{}XUV}}{4\pi{}d^2}\frac{S_{\mathrm{XUV}}}{\Phi_{0}K_{\mathrm{tide}}},
\end{equation}

where $L_{\rm{}XUV}$ is the radiative energy input (X-ray and extreme UV stellar flux) and $S_{\rm{}XUV}=\pi{}R_{\rm XUV}^2$ is the cross-section surface that collects the high-energy irradiation, where $R_{\rm XUV}$ is the XUV radius (i.e., the radius where most of the XUV flux is absorbed). Furthermore, $\Phi_0=GM_{\rm pl}/R_{\rm pl}$ is the gravitational potential of the planet, where the planetary radius $R_{\rm pl}$ is defined as the optical photosphere radius and $K_{\rm tide}$ is a correction factor accounting for atmospheric loss enhancement due to the action of tidal forces because the planetary Roche lobe can be close to the planet's surface \citep{Erkaev2007}. Its expression is given by :

\begin{equation}
K_{\rm{tide}}=1-\frac{3}{2\xi}+\frac{1}{2\xi^3}
\end{equation}

\begin{equation}
\label{eq:xi}
\xi=\frac{R_{\rm{}Roche}}{R_{\rm{}pl}}=\left(\frac{M_{\mathrm{pl}}}{3M_{\mathrm{s}}}\right)^{1/3}\frac{a}{R_{\rm{}pl}}\left(1+\frac{e^2}{2}\right).
\end{equation}

To account for the effect of a noncircular orbit, Eq. (\ref{eq:xi}) uses the average orbital distance \citep{Guo2010} instead of the semi-major axis alone \citep{Erkaev2007}. $\Phi_{0}K_{\rm tide}$ comes into play in Eq. (\ref{eq:massloss}) because it represents the gravitational potential difference between the planetary radius and the Roche-lobe radius for a planet affected by stellar tidal forces \citep{Erkaev2007}. We calculate this potential using the optical radius $R_{\rm pl}$  \citep[as, e.g.,][]{Watson1981,Lammer2003,Erkaev2007,Salz2016a,Kubyshkina2018a} rather than the XUV radius \citep[as, e.g.,][]{Baraffe2004,Lopez2012,Kurokawa2014,Lopez2017} to account for the energy required to lift material from $R_{\rm pl}$ to $R_{\rm XUV}$ \citep{Salz2016a}.

In contrast to \citet{Barnes2020}, we differentiate between $R_{\rm pl}$ and $R_{\rm XUV}$, which can otherwise lead to relatively high inaccuracies in the derived mass-loss rate \citep{Salz2016a}. The most direct way to define the XUV radius at first order is to add a characteristic height to the planetary radius \citep[e.g.,][]{Lopez2017}. In the JADE code though, we chose to define the XUV radius via a more accurate analytical relation \citep{Salz2016a}:

\begin{multline}
\label{eq:RXUV}
\log_{10}(R_{\rm{}XUV}/R_{\rm{}pl})=\max(0.0, \\
-0.185\log_{10}(GM_{\rm{}pl}/R_{\rm{}pl})+0.021\log_{10}(F_{\rm{}XUV})+2.42),
\end{multline} 

where $R_{\rm XUV}$ depends on the gravitational potential of the planet and the stellar XUV flux $F_{\mathrm{XUV}}$ (all the quantities are in CGS). $\epsilon$ in Eq. (\ref{eq:massloss}) is the efficiency of photo-evaporation (fraction of the energy input that is available for atmospheric heating). It is usually taken to be between 0.1 and 0.2 \citep[e.g.,][]{Jackson2010,Valencia2010,Lopez2012,Shematovich2014,Jin2014,Salz2016b,Lopez2017}. In order to be consistent with our definition of $R_{\rm XUV}$, we chose to define $\epsilon$ by using the analytical formula of \citet{Salz2016a}:

\begin{equation}
\label{eq:epsilon}
\log_{10}\epsilon=
\begin{cases}
-0.50-0.44(v-12.00) & \text{if } v\leqslant{}13.11, \\
-0.98-7.29(v-13.11) & \text{if } v>13.11,
\end{cases}
\end{equation}

where $v=\log_{10}(GM_{\rm pl}/R_{\rm pl})$. Equations (\ref{eq:RXUV}) and (\ref{eq:epsilon}) were derived from self-consistent simulations of upper atmospheric structures over a large range of stellar irradiations, planetary masses, and optical radii. Using these equations allows us to cover a wider range of evaporation regimes (e.g., recombination-limited, X- or EUV-driven) than in the naive energy-limited approach.

Finally, $d$ in Eq. (\ref{eq:massloss}) is the planet-star distance. As this quantity varies over one orbit, we averaged Eq. (\ref{eq:massloss}) over the inner orbital period in order to make it compatible with the dynamical secular equations (Appendix \ref{app:avg}). This step is crucial since it ensures one of the novelties of the JADE code: coupling the possible complex dynamical evolution of the system with photo-evaporation. The secularized mass-loss rate after all possible simplifications is then:
\begin{equation}
\label{eq:masslosssec}
\Big\langle{\dot{M}_{\rm{env}}}\Big\rangle=\epsilon\frac{L_{\rm{XUV}}R_{\rm{}pl}R_{\rm{}XUV}^2}{4GM_{\rm{}pl}K_{\rm{tide}}\sqrt{1-e^2}a^2}.
\end{equation}

One possible caveat is the assumption that the atmosphere reacts instantly to variations in the incoming stellar flux. It has been suggested, for example, that stellar flares have a limited impact on mass loss due to the delayed response of the upper atmosphere to the sharp and short increase in flux \citep{Chadney2017, Bisikalo2018}. In the present framework, the incoming flux can vary rapidly in the vicinity of the periastron during high-eccentricity phases, but such variations remain more gradual than a stellar flare.

\subsubsection{Evolution of the stellar luminosity}

The stellar luminosity is an important quantity in our model. It intervenes several times in the equations governing the planet's atmospheric structure and thus its evolution. We separate this luminosity into 3 main spectral bands: bolometric $L_{\rm bol}$, X-rays $L_{\rm X}$, and extreme ultra-violet $L_{\rm EUV}$. $L_{\rm bol}$ is used to determine the equilibrium temperature of the planet, while the XUV luminosity $L_{\rm XUV}=L_{\rm X}+L_{\rm EUV}$ is the cornerstone of the atmospheric mass-loss equation.

The preferred solution we implemented in the JADE code is to provide time-tabulated luminosities, which are then interpolated at each time step. This approach allows to directly use the results of relevant stellar simulations. In the event those tables cannot be provided, one can use the analytic expressions implemented in the JADE code. In this case, the bolometric luminosity stays constant in time, as its variation is usually negligible compared to the variation of the X and EUV luminosities \citep[as with the Sun,][]{Guinan2009}. We derive $L_{\rm X}$ from $L_{\rm bol}$ using the model of \citet{Jackson2012}, in which the ratio $L_{\rm X}/L_{\rm bol}$ evolves in two stages: a first phase where it remains constant and a second one, when the star is older than a saturation age, where it drops as a decreasing power-law \citep{Jackson2012}:

\begin{align}
\label{eq:Lx}
L_{\rm{}X}/L_{\rm{}bol}=
\begin{cases}
(L_{\rm{}X}/L_{\rm{}bol})_{\rm{}sat} & \text{if } t\leqslant{}\tau_{\rm{}sat}, \\
(L_{\rm{}X}/L_{\rm{}bol})_{\rm{}sat}(t/\tau_{\rm{}sat})^{-\alpha} & \text{if } t>\tau_{\rm{}sat},
\end{cases}
\end{align} 

where $(L_{\rm{}X})_{\rm{}sat}$, $\tau_{\rm sat}$, and $\alpha$ are constant parameters that can be imposed or automatically set to the average values from \citet{Jackson2012}. Finally, we derive $L_{\rm EUV}$ directly from $L_{\rm X}$ using a single power-law \citep{King2018}:

\begin{equation}
L_{\rm{}EUV}/L_{\rm{}X}=\alpha\left(\frac{L_{\rm{}X}}{4\pi{}a^2}\right)^\gamma,
\end{equation}

where $\alpha$ and $\gamma$ are constant parameters set to the values given by \citet{King2018} for a boundary at 100 $\AA$ between X-ray and EUV.

\begin{figure*}
\centering
\includegraphics[width=8.5cm]{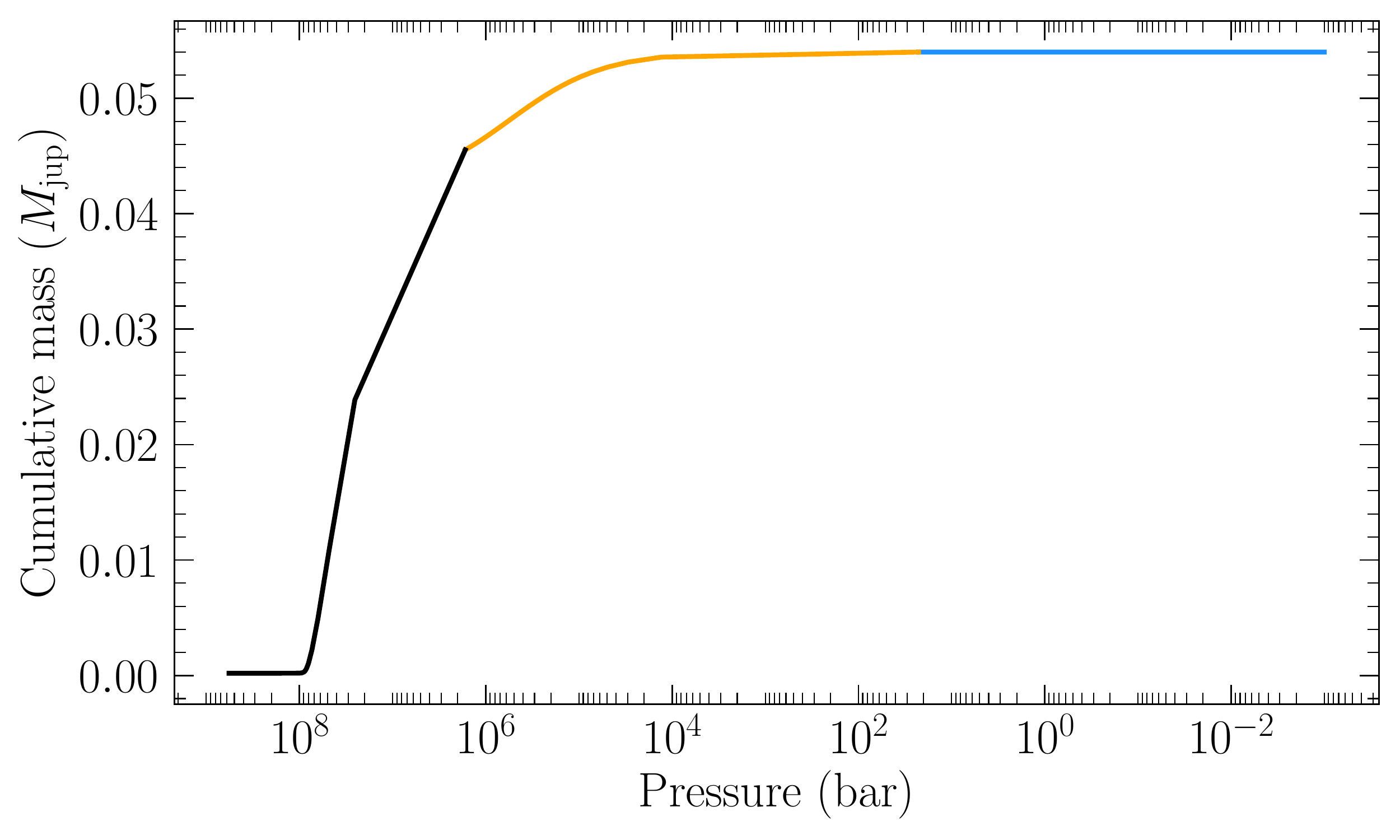}
\includegraphics[width=8.5cm]{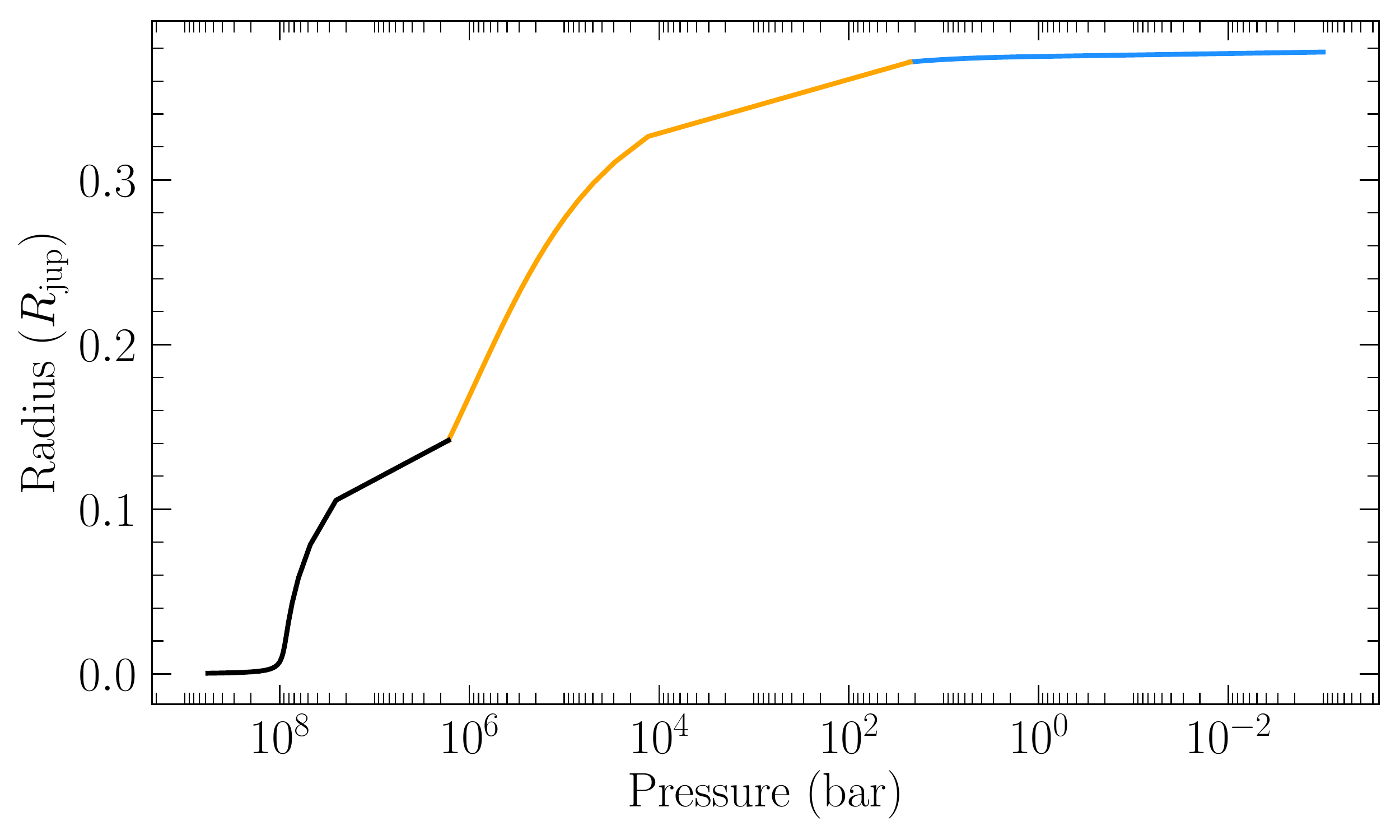}
\includegraphics[width=8.5cm]{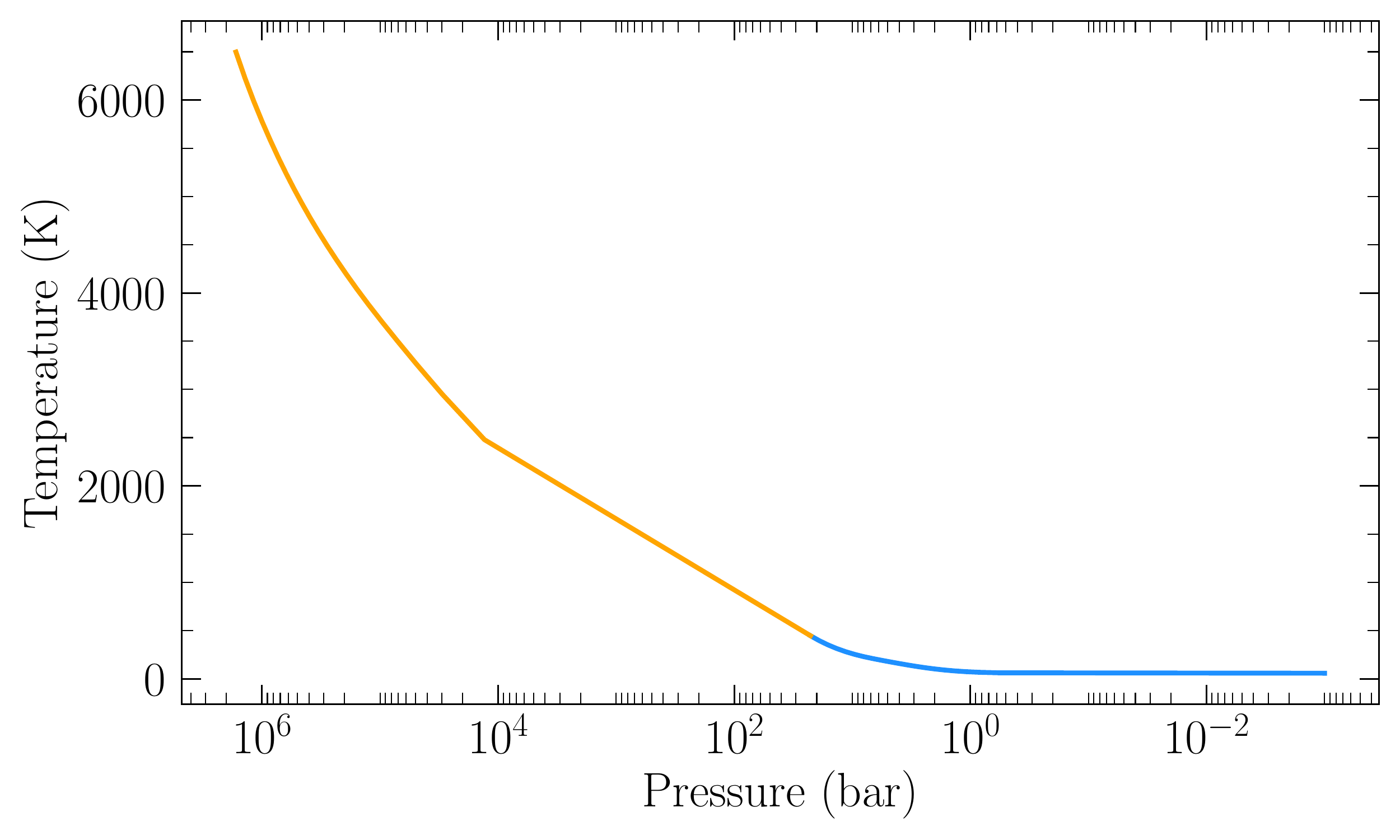}
\includegraphics[width=8.5cm]{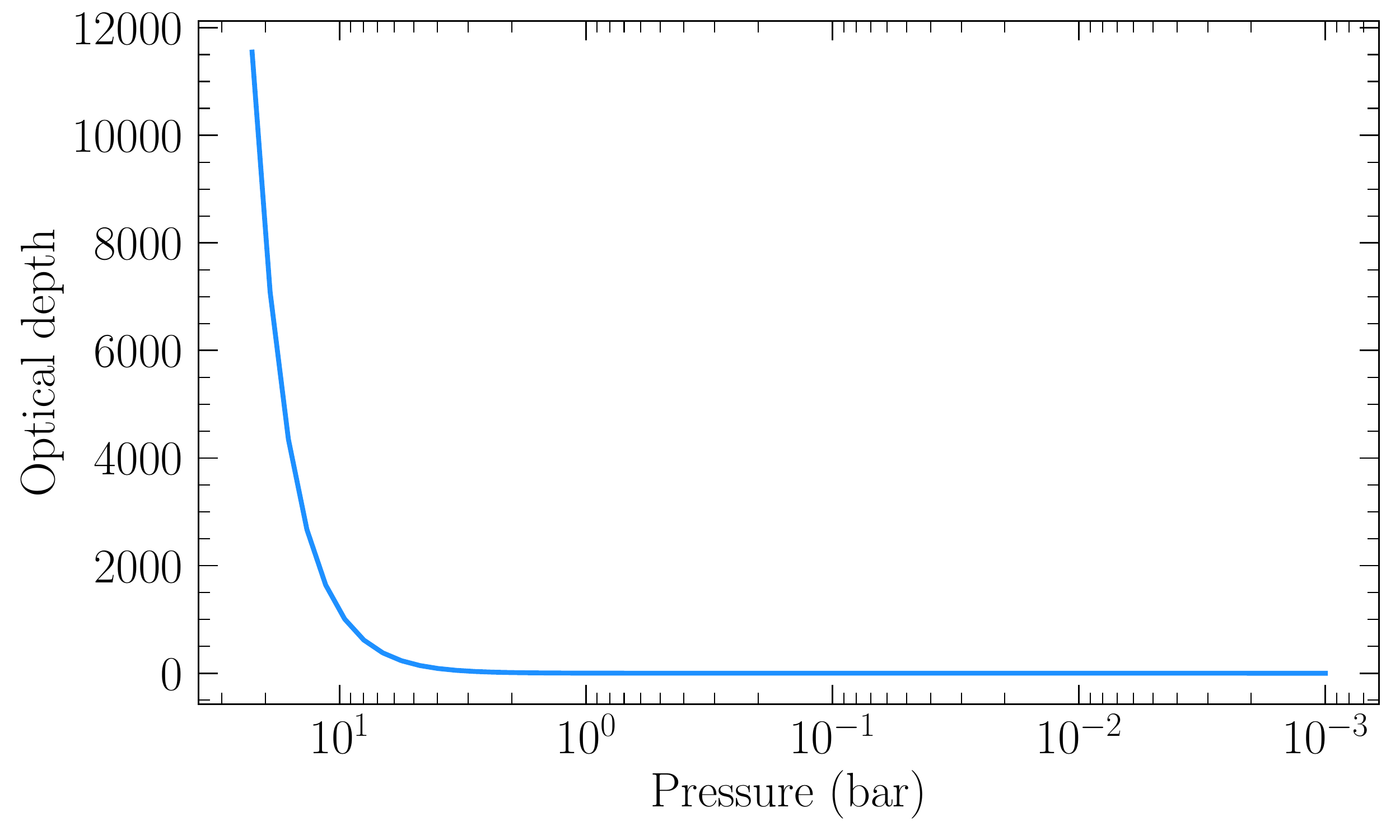}
\caption{Profiles for the test simulation of Neptune as a function of $\log{P}$. The blue curves correspond to Zone A, the orange curves correspond to Zone B, and the black curves to the rocky core. The parameters of the simulation are: $a=30$ AU, $M_{\rm core}=0.85M_{\rm pl}$, $M_{\rm pl}=1M_{\rm nept}$, $L_{\rm bol}=1L_\odot$, and $Y_{\rm He}=0.2$. \textit{Top left}: cumulative mass of the planet. \textit{Top right}: radius of the planet. \textit{Bottom left}: temperature. Bottom right: optical depth in thermal wavelengths.}
\label{fig:atmo}
\end{figure*}

\begin{figure}
\resizebox{\hsize}{!}{\includegraphics{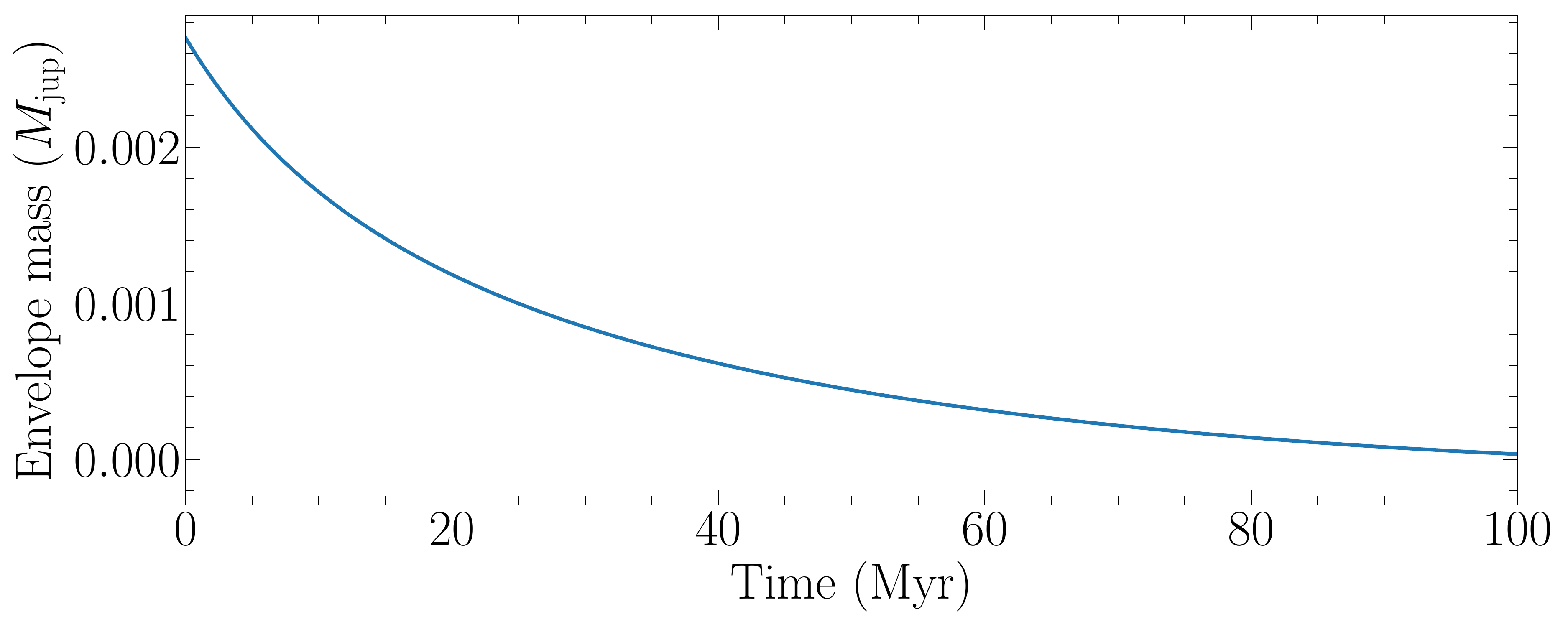}}
\resizebox{\hsize}{!}{\includegraphics{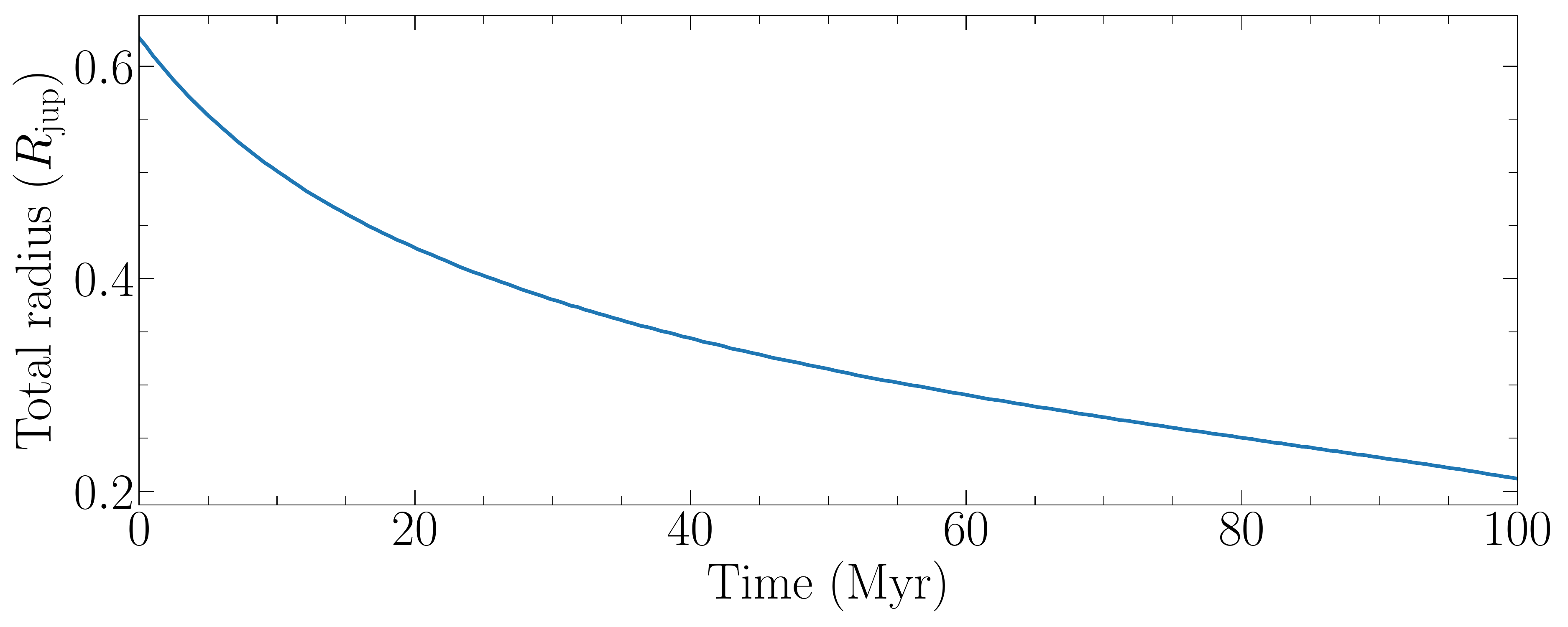}}
\caption{Secular evolution of the test system over 100 Myr. The only active mechanism is photo-evaporation. The parameters of the simulation are: $M_{\rm pl}=1M_{\rm nept}$, $M_{\rm core}=0.95M_{\rm pl}$, $L_{\rm bol}=1L_\odot$, $Y_{\rm He}=0.2$, and $a=0.05$ AU. There is no perturber in this simulation. The envelope has completely evaporated at the end of the simulation. \textit{Top}: envelope's mass. \textit{Bottom}: planetary radius.}
\label{fig:atmoevap}
\end{figure}

\subsubsection{Validation of the atmospheric model}

Two different validations have to be conducted. The first one is related to the static part, that is to say the atmospheric structure integration (Sect. \ref{sect:atmostruc}). The second one is related to the evolutive part (photo-evaporation, Sect. \ref{sect:evap}).

So as to validate the JADE code's atmospheric structure integration, we sought to recover the radius of Neptune in the solar system from its known parameters: $a=30$ AU, $Y_{\rm He}=0.2$ \citep{Hubbard1995}, and $M_{\rm env}/M_{\rm pl}=0.15$ \citep{Nettelmann2013,Frelikh2017}. We use $L_{\rm bol}=1L_\odot$. $L_{\rm XUV}$  and $L_{\rm pl}$ are automatically computed using the procedure presented in the previous sections. We successfully find $R_{\rm pl}=0.370R_{\rm jup}\simeq1.05R_{\rm nept}$ after 5 Gyr. The present mass, radius, temperature, and optical depth profiles as a function of pressure are plotted in Fig. \ref{fig:atmo}. These profiles show the expected behaviors. The deeper parts of the envelope are much hotter and more compressed than the surface. The shape of the $T-P$ profile corresponds to what is seen in literature \citep[e.g.,][]{Fortney2008,Mordasini2012,Jin2014}. Particularly, the small change of the temperature in Zone A is due to the nongray effects. The slope variation of the temperature in Zone B is due to the transition from a radiative to a convective regime. The integration correctly stops at the right value of $\tau$ in Zone A and of $m$ in Zone B.

To test photo-evaporation's effect, we simulated the atmospheric mass loss of a $M_{\rm pl}=0.054M_{\rm jup}\simeq 1M_{\rm nept}$ Neptune with 5\% of gaseous envelope. We set the value of $a$ to 0.05 AU (orbital period of about 10 days) to better see the effect of photo-evaporation. All the dynamical features are switched off, photo-evaporation is the only active mechanism. The results are presented in Fig. \ref{fig:atmoevap}. The envelope has been completely stripped away in approximately 100 Myr, mass and radius dropping until they reach the rocky core.


\section{Application to the GJ436 system}
\label{sect:gj436}

\begin{figure*}
\centering
\includegraphics[width=8.5cm]{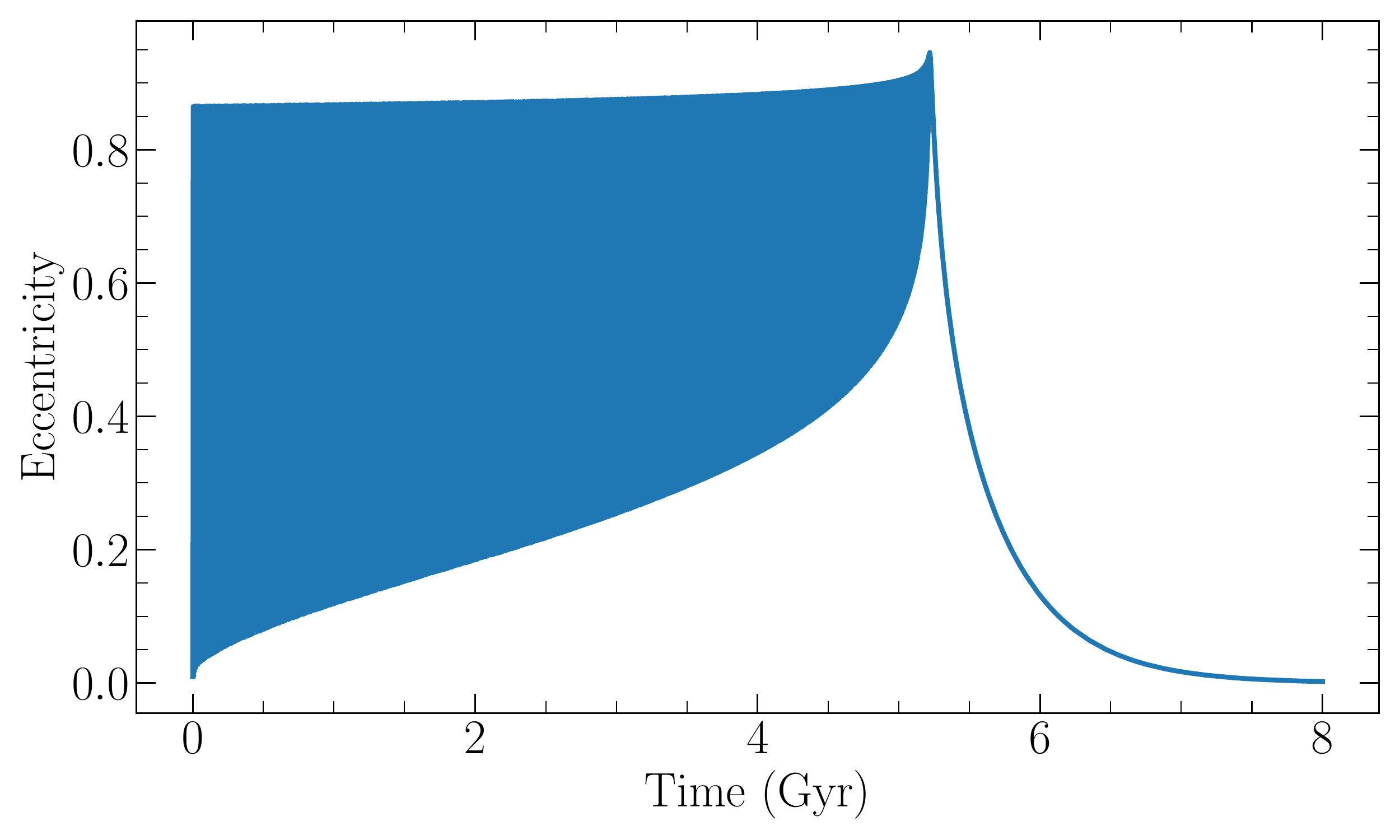}
\includegraphics[width=8.5cm]{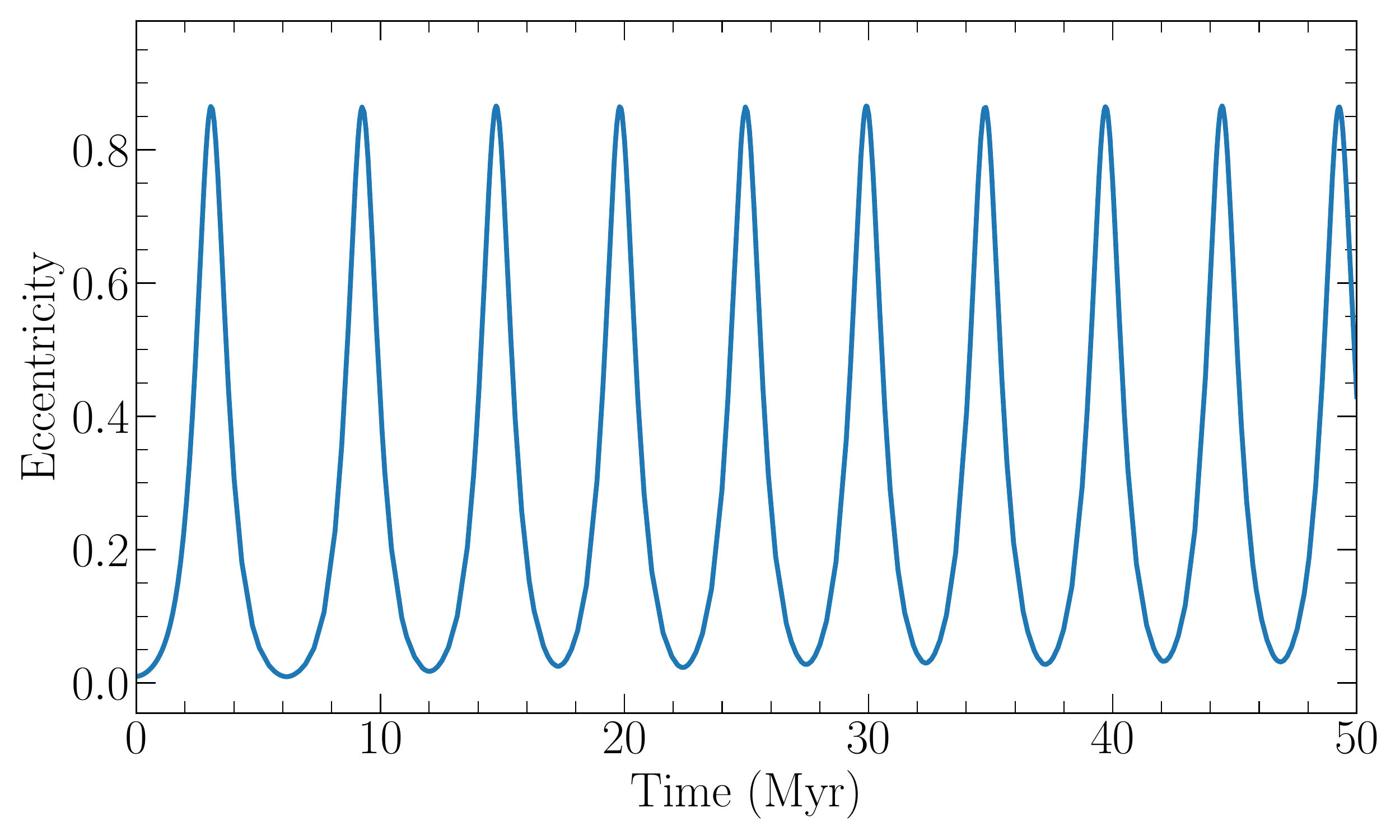}
\includegraphics[width=8.5cm]{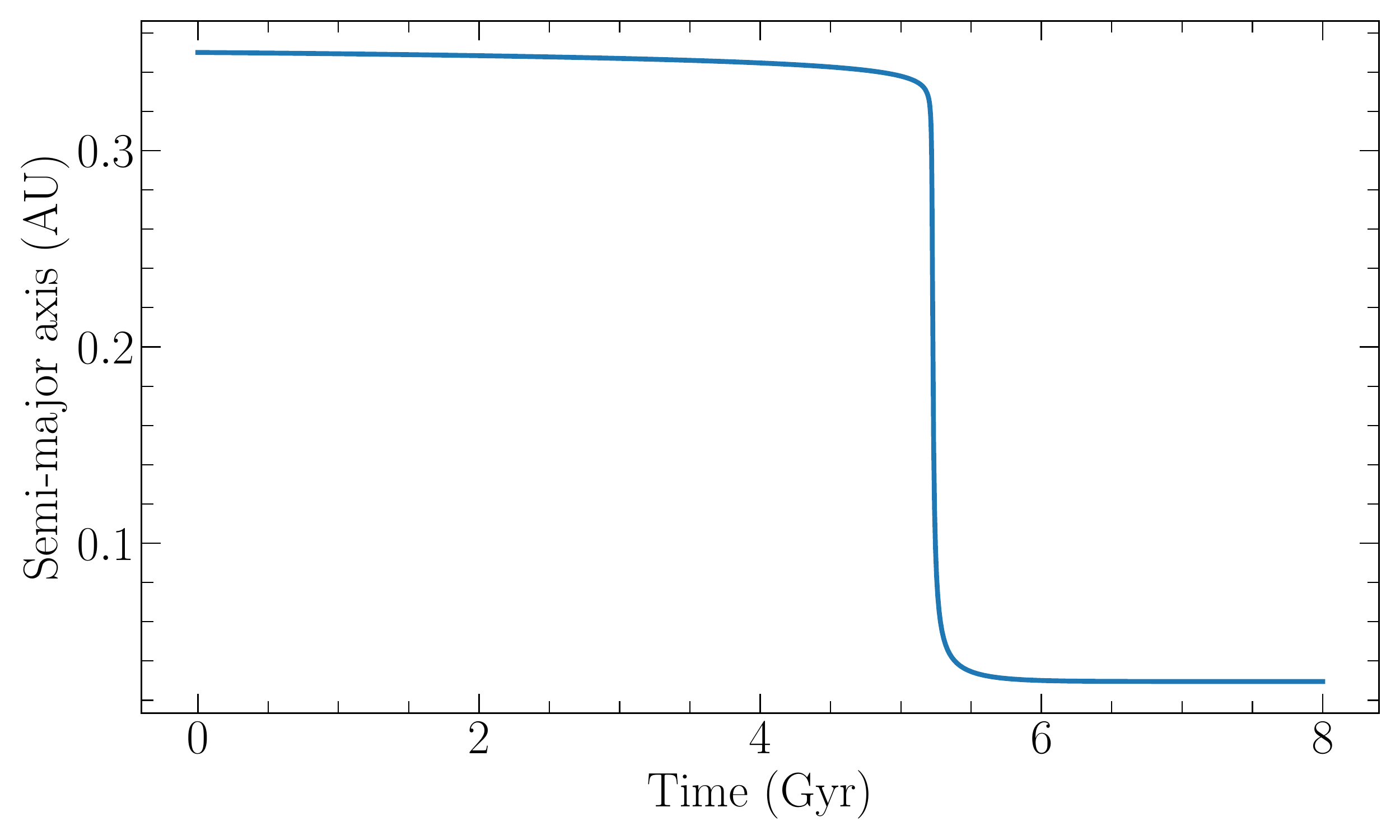}
\includegraphics[width=8.5cm]{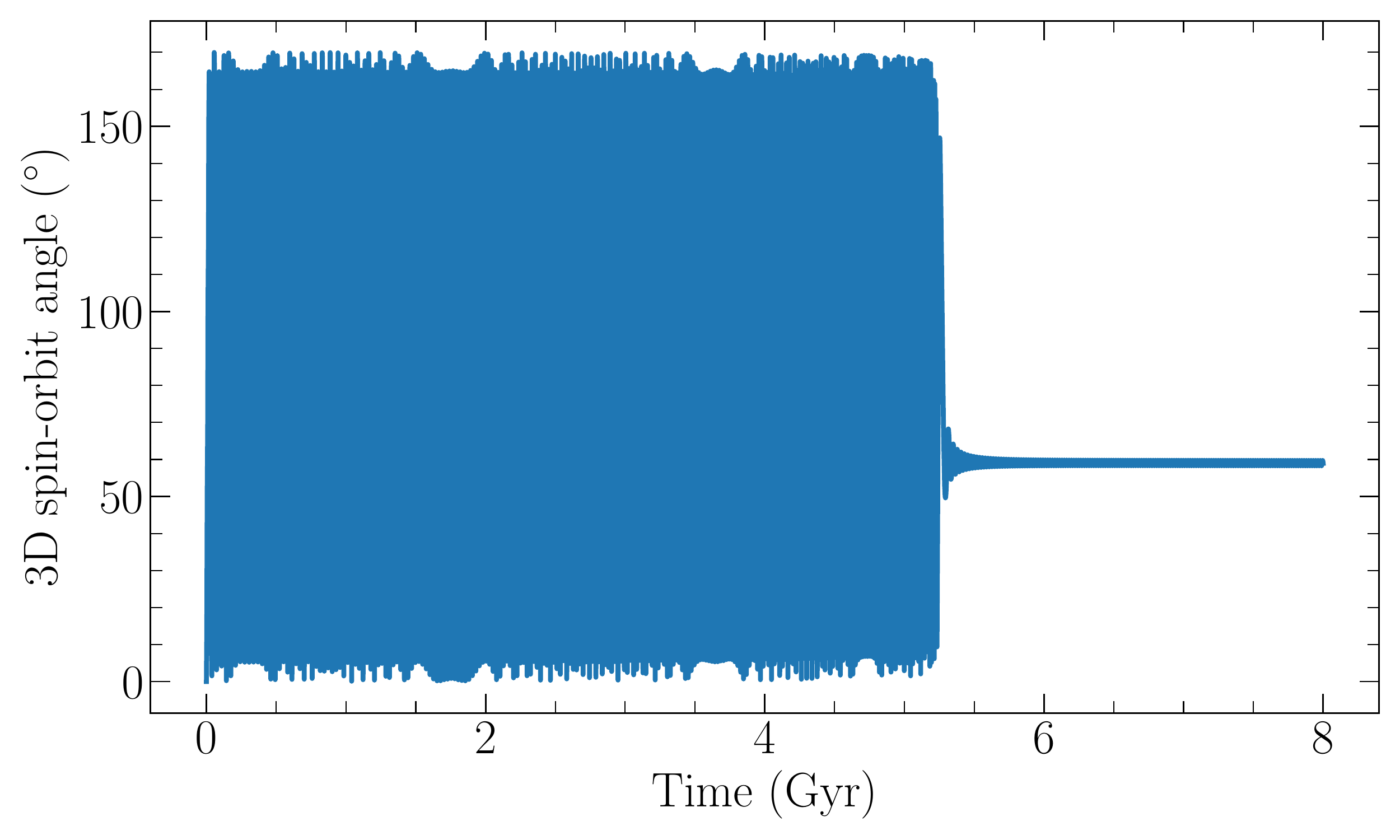}
\caption{Secular evolution of the GJ436 system. The simulation was performed including only the dynamical features. See text for initial conditions. \textit{Top left}: eccentricity on a long time scale of 8 Gyr where the two-part behavior is clear. \textit{Top right}: eccentricity on an intermediate time scale of 50 Myr where the Kozai-Lidov oscillations are apparent. \textit{Bottom left}: semi-major axis on secular time scale. \textit{Bottom right}: 3D spin-orbit angle on secular time scale.}
\label{fig:GJ436_dyn}
\end{figure*}

The GJ436 system is a prime candidate to test the JADE code. The M-dwarf star is known to host a transiting close-in Neptune-mass planet, GJ436 b \citep{Butler2004,Gillon2007}, which has been the subject of growing interest in recent years. The monitoring of its transits coupled to RV measurements constrained its bulk and orbital parameters. GJ436 b has a significant nonzero eccentricity, $e=0.14 \pm 0.01$, despite a small orbital period, $P \simeq 2.64$ days \citep{Lanotte2014}, which should have led tidal forces to circularize the orbit a long time ago. Several theories were proposed to solve this riddling case. The most straightforward one is the possible weakness of tidal forces \citep[e.g.,][]{Mardling2008}. However, this argument is mainly related to our poor knowledge of the tidal dissipation factor $Q_{\rm pl}$, which needs to be set at 10 times the upper-bound estimate for a Neptune to explain the eccentricity of GJ436 b. Among the other possible scenarios, \citet{Beust2012} proposed an explanation to its high eccentricity based on a Kozai resonance induced by a distant perturber. This scenario is supported by \citet{Bourrier2018a}, as it can explain both their measurement of GJ436 b misaligned orbit ($\lambda \simeq 72^\circ$) and its high eccentricity. These authors further suggested that this mechanism could explain the strong atmospheric escape of GJ436 b \citep{Ehrenreich2015,Lavie2017} despite its presence at the fringes of the desert more than 4 Gyr after the planet's formation, as evaporation would have been triggered by its late Kozai migration.

While \citet{Bourrier2018a} were able to explain the observed orbital architecture of GJ436 b with simulations of Kozai migration, they did not couple the dynamical evolution of the system with the evaporation of the planet. The interest of studying this coupling is enhanced by the possibility to constrain the coupled evolution with measurements of both the planet's orbital architecture and its mass-loss rate \citep{Bourrier2015c,Bourrier2016}. This is where the added value of the JADE code comes into play.


\subsection{Pure Kozai migration}
\label{sect:gjkozai}

As the secular evolution of GJ436 under Kozai resonance has already been investigated, it can be used to benchmark the dynamical part of the JADE code. As pointed out by \citet{Beust2012} and \citet{Bourrier2018a}, its orbit should already have been circularized assuming $Q_{\rm pl}=10^5$ and the current age of the system \citep[4 - 8 Gyr,][]{Bourrier2018a}. But the perturbing effect of the companion delays the circularization. The stellar parameters are $M_{\rm s}=0.445 M_\odot$ and $R_{\rm s}=0.449 R_\odot$ \citep{Mann2015}. We assume the presence of a distant perturber, GJ436 c, with the following properties: $M_{\rm pert}=0.1 M_{\rm jup}$, $a_{\rm pert}=5.8$ AU, and $e_{\rm pert}=0.03$ to ensure a Kozai resonance compatible with the age of the system \citep[see Fig. 4 of][]{Bourrier2018a}. We carry out a purely dynamical simulation (the atmospheric structure and evolution are not considered) including tides, general relativity, and the action of the perturber. The bulk properties of GJ436 b do not evolve over time and are set to $M_{\rm pl}=0.0799 M_{\rm jup}$ and $R_{\rm pl}=0.374 R_{\rm jup}$ \citep{Bourrier2018a}. If GJ436 b had migrated early-on to its present location, tides and relativistic effects would have been strong enough to inhibit all Kozai cycles \citep{Beust2012}. As in \citet{Beust2012,Bourrier2018a}, we assume GJ436 b formed about ten times further ($a=0.35$ AU) than its current orbit \citep[$a=0.0308$ AU,][]{Lanotte2014} and set $i_{\rm mut}=85^\circ$ so as to generate a strong Kozai resonance. The initial inner eccentricity is set close to zero ($e=0.01$) and an initially null spin-orbit angle is assumed. It should be noted that we do not address the origin of the high initial mutual inclinations we use. Three-body disk-driven resonance \citep{Petrovich2020} provides interesting leads on this issue. While this mechanism does not explain the residual eccentricities of billion-year-old close-in planets and the orbital architectures of the most massive planets, it could explain how Neptune-mass and smaller planets acquire eccentric orbits, inclined with respect to their companion and misaligned at the end of the disk phase.

\begin{figure*}
\centering
\includegraphics[width=8.5cm]{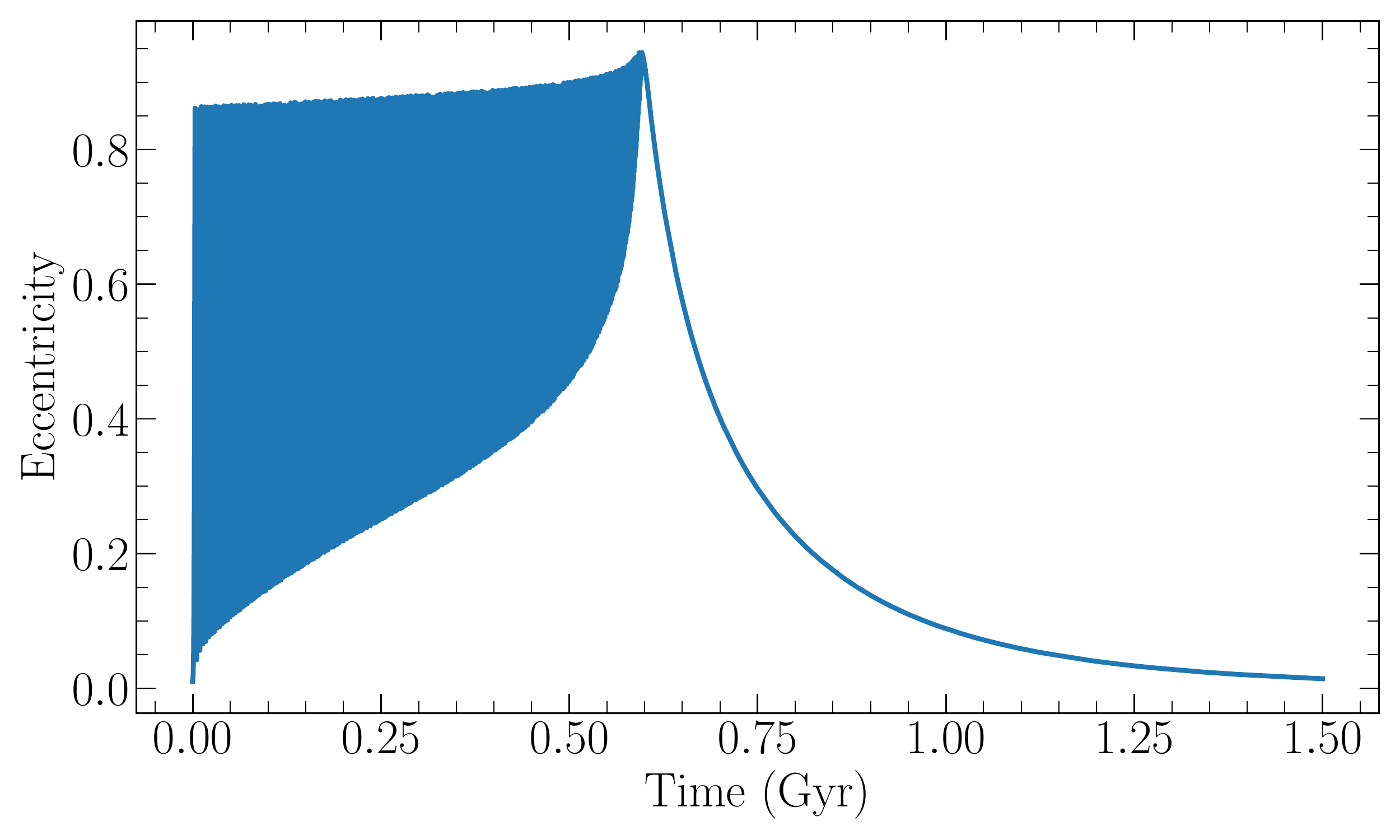}
\includegraphics[width=8.5cm]{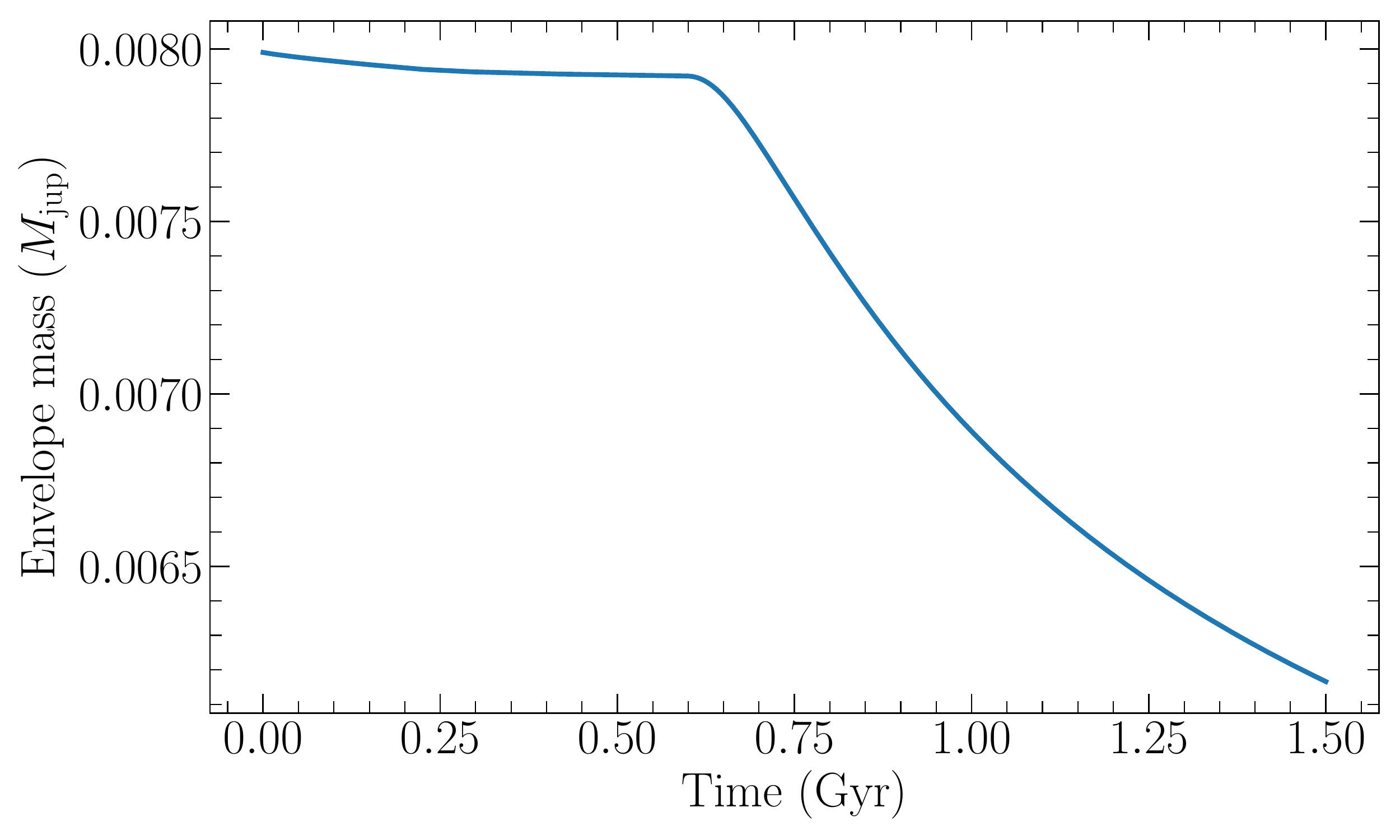}
\includegraphics[width=8.5cm]{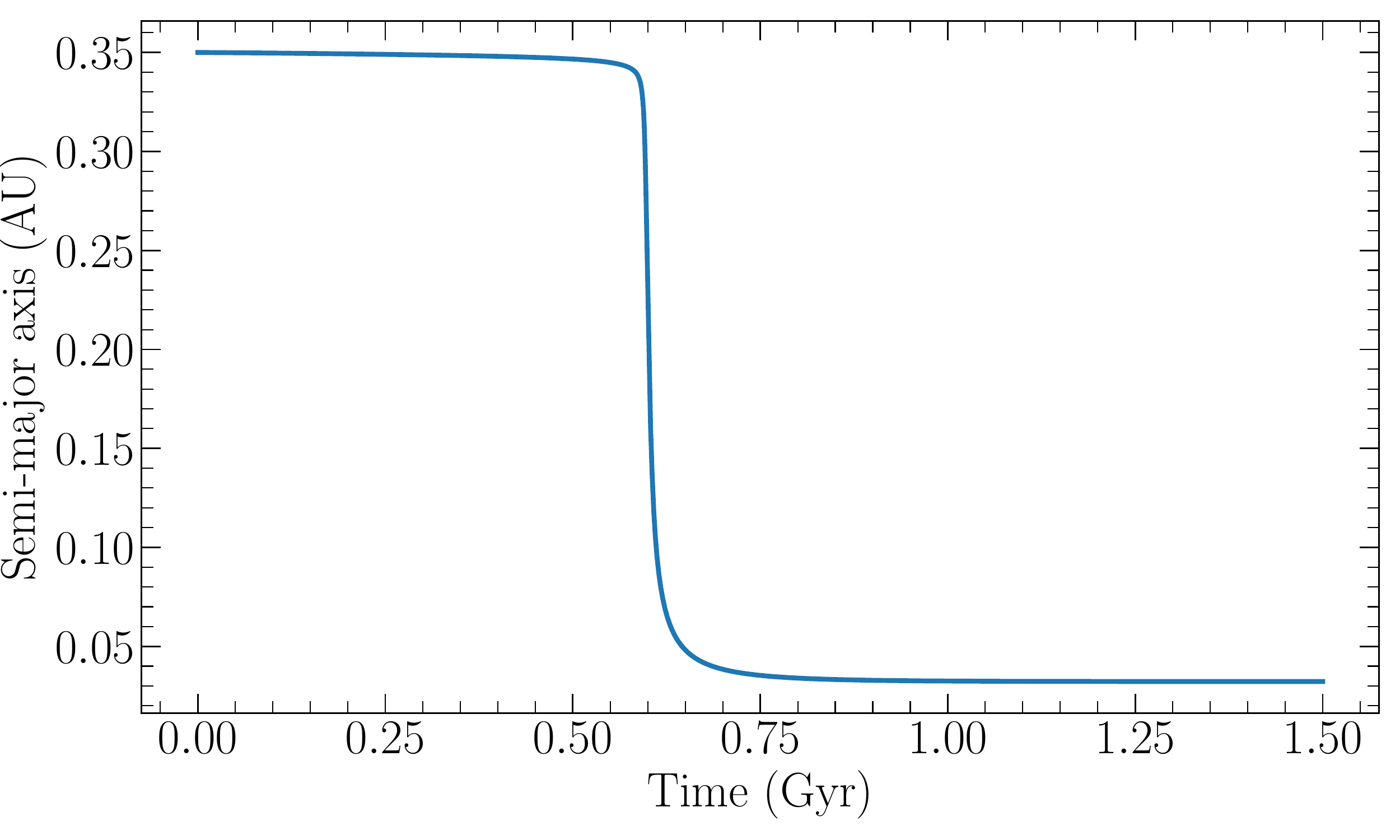}
\includegraphics[width=8.5cm]{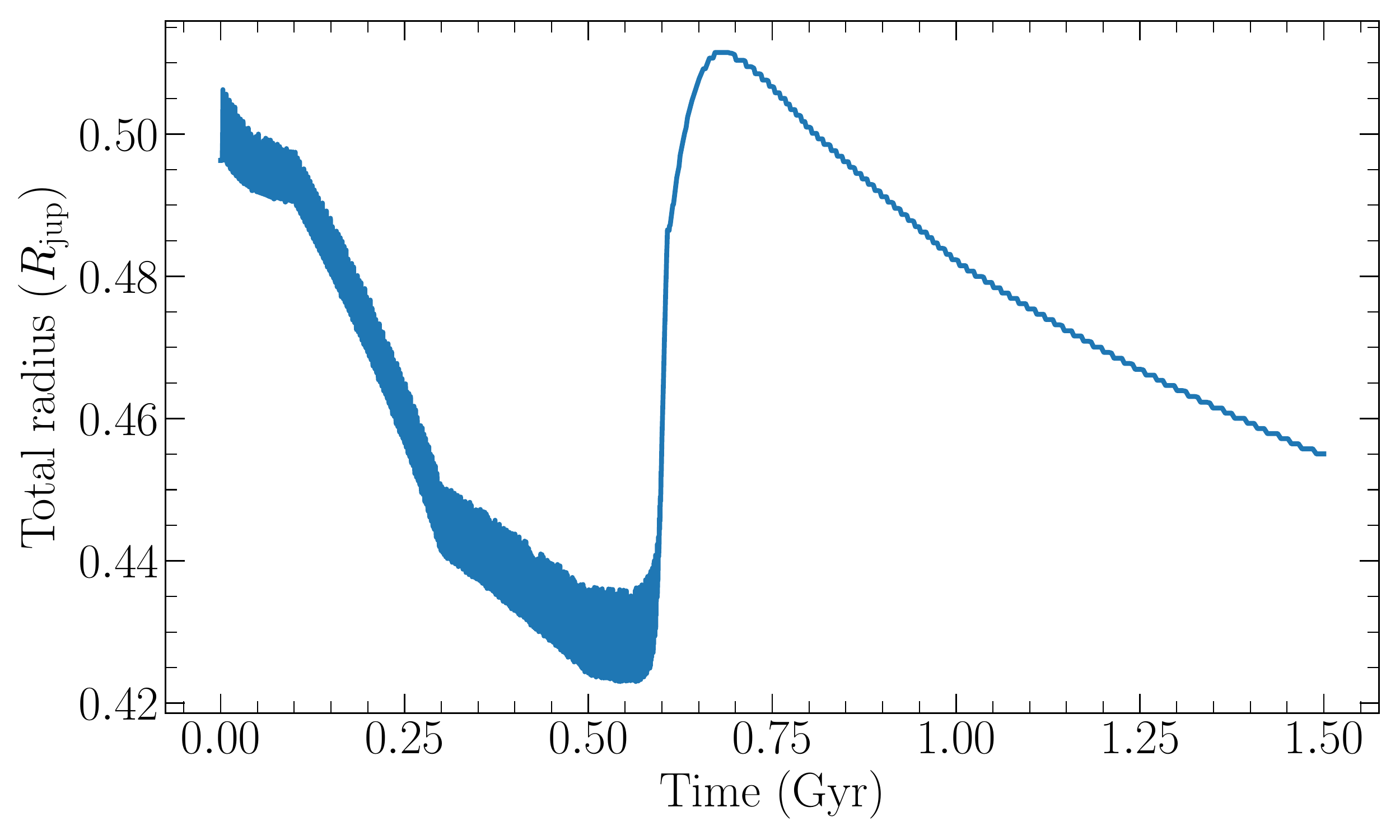}
\caption{Secular evolution of the GJ436 system. The simulation was performed including both dynamical features and an atmospheric structure. Photo-evaporation is active. See text for initial conditions. \textit{Top left}: eccentricity as a function of time. \textit{Top right}: mass of the envelope as a function of time. \textit{Bottom left}: semi-major axis as a function of time. \textit{Bottom right}: planetary radius as a function of time.}
\label{fig:GJ436_atmo}
\end{figure*}

The results are illustrated in Fig. \ref{fig:GJ436_dyn} on two different time scales. The second subplot presents an intermediate time scale of 50 Myr to showcase the Kozai-Lidov eccentricity oscillations. Equation (\ref{eq:taukozai}) yields a characteristic Kozai period of $\tau_{\rm Kozai}=1.4$ Myr, in rough agreement with Fig. \ref{fig:GJ436_dyn}. The other subplots highlight the dynamical evolution on a secular time scale of 8 Gyr that is comparable to the age of the system. A clear two-part behavior is observed. The first phase shows very fast Kozai oscillations compared to the time scale on which the simulation is performed. This resonance is wrapped by an envelope that substantially shrinks over time. The first phase comes to an end after a characteristic transition time scale $\tau_{\rm trans}$ (a bit higher than 5 Gyr) when the bottom eccentricity of this envelope reaches the same value as the top eccentricity. After that, during the second phase of the evolution, the planet undergoes tidal damping as the orbit gets gradually circularized, which translates into the sharp decrease in the semi-major axis and slower decrease in the eccentricity. Moreover, the spin-orbit angle widely oscillates during the first phase and comes out of resonance with a high value, which later continues to weakly oscillate during the second phase. Hence, even if the orbit was initially aligned, a strong Kozai resonance can substantially tilt it, providing a possible origin for the high measured obliquity today.

We thus satisfyingly find the same results as \citet{Beust2012,Bourrier2018a}, which validate our dynamical integrator and confirm Kozai cycles as a possible explanation to GJ436 b misaligned and eccentric orbit. By making the planet form ten times further than its current position, it spends several Gyr in the Kozai resonance and emerges with a high obliquity and eccentricity, recently enough that its orbit would not have circularized yet.


\subsection{Kozai migration and evaporating atmosphere}

In this section, we carry out the same simulation of GJ436 b as in Sect. \ref{sect:gjkozai} but we now account for the presence of an evaporating atmosphere. To perform this, we consider that 10\% of the planet's mass is made up of a H/He atmosphere characterized by a $Y_{\rm He}=0.2$ helium fraction, similarly to Neptune. The temporal evolution of the stellar bolometric and XUV luminosities is dictated by tabular values that were derived using dedicated stellar simulations of GJ436 and that will be detailed in a follow-up paper. All the features contributing to the planetary internal energy (Sect. \ref{sect:atmostruc}) are taken into account. 

\begin{figure}
\resizebox{\hsize}{!}{\includegraphics{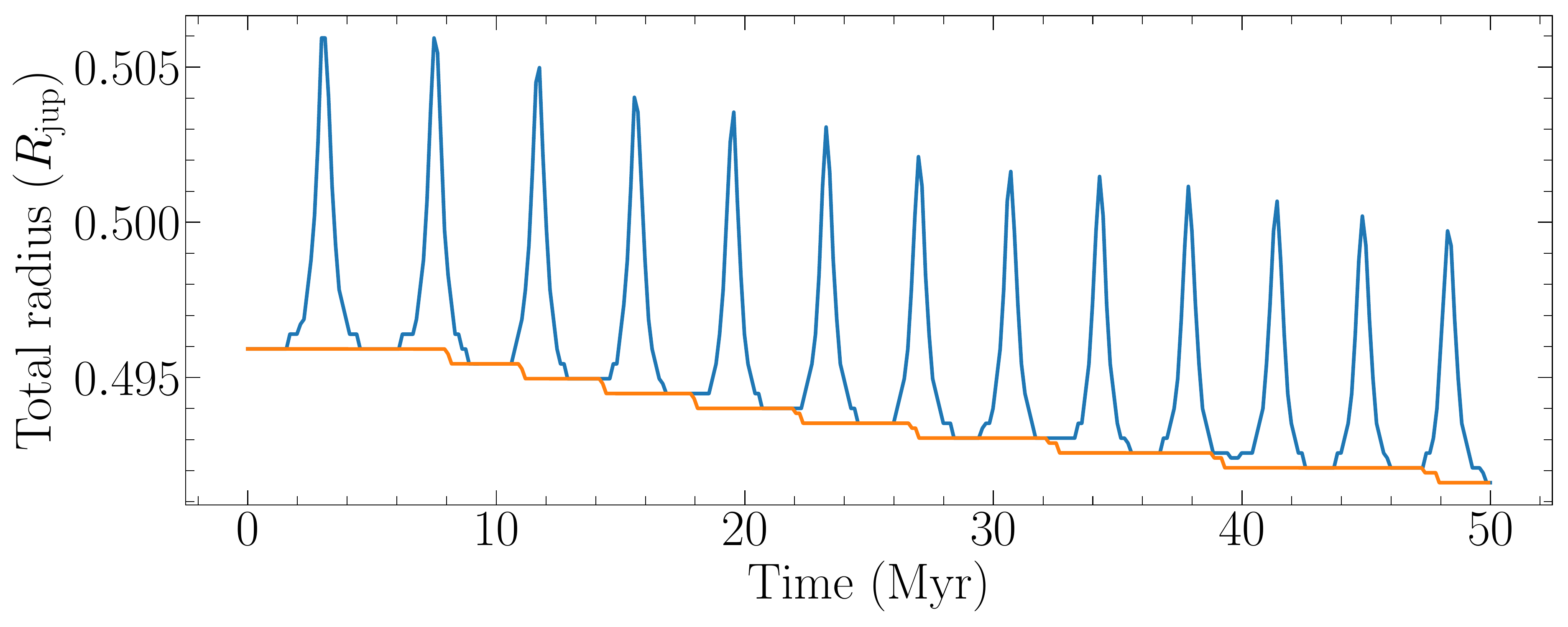}}
\resizebox{\hsize}{!}{\includegraphics{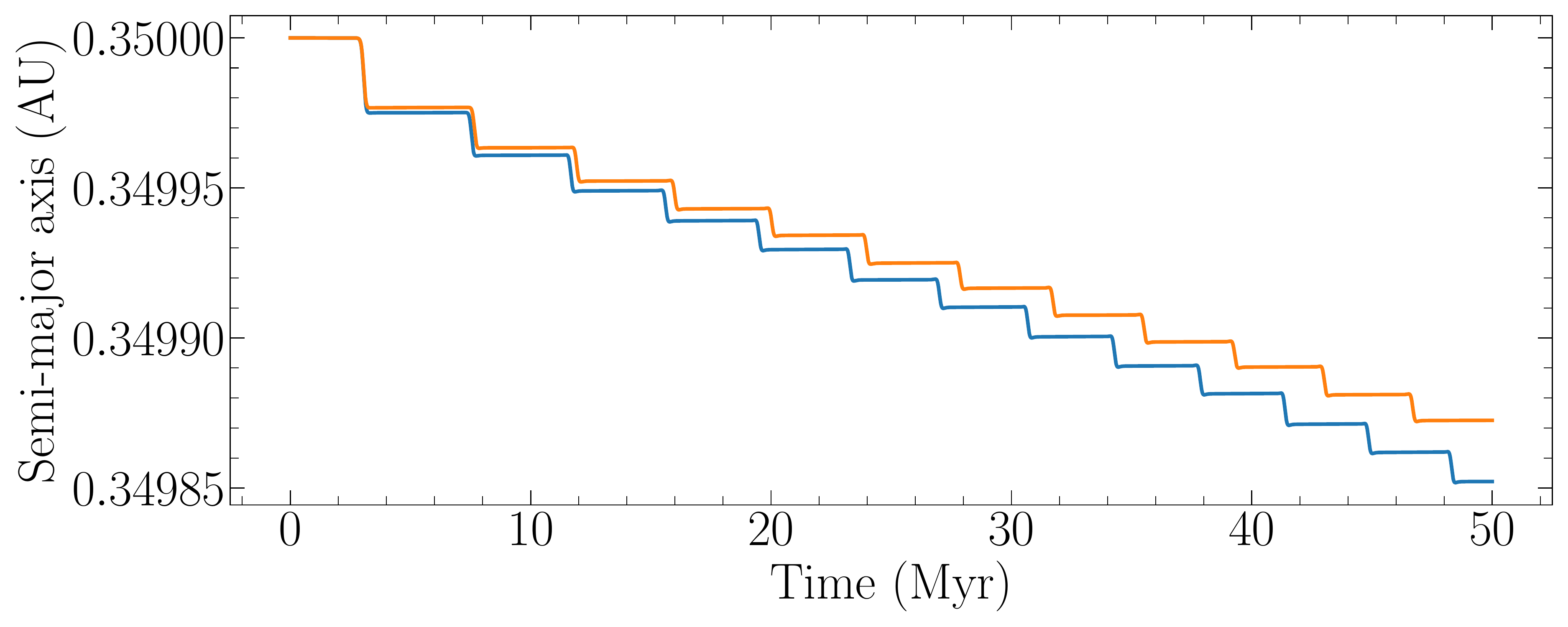}}
\caption{\textit{Top}: evolution of the planetary radius during the first 50 Myr. \textit{Bottom}: evolution of the semi-major axis during the first 50 Myr. The blue lines represent the simulation where $L_{\rm bol}$ is averaged over the inner orbit, as opposed to the orange lines where it is not.}
\label{fig:GJ436_atmo_sec}
\end{figure}

\begin{figure*}
\centering
\includegraphics[width=8.5cm]{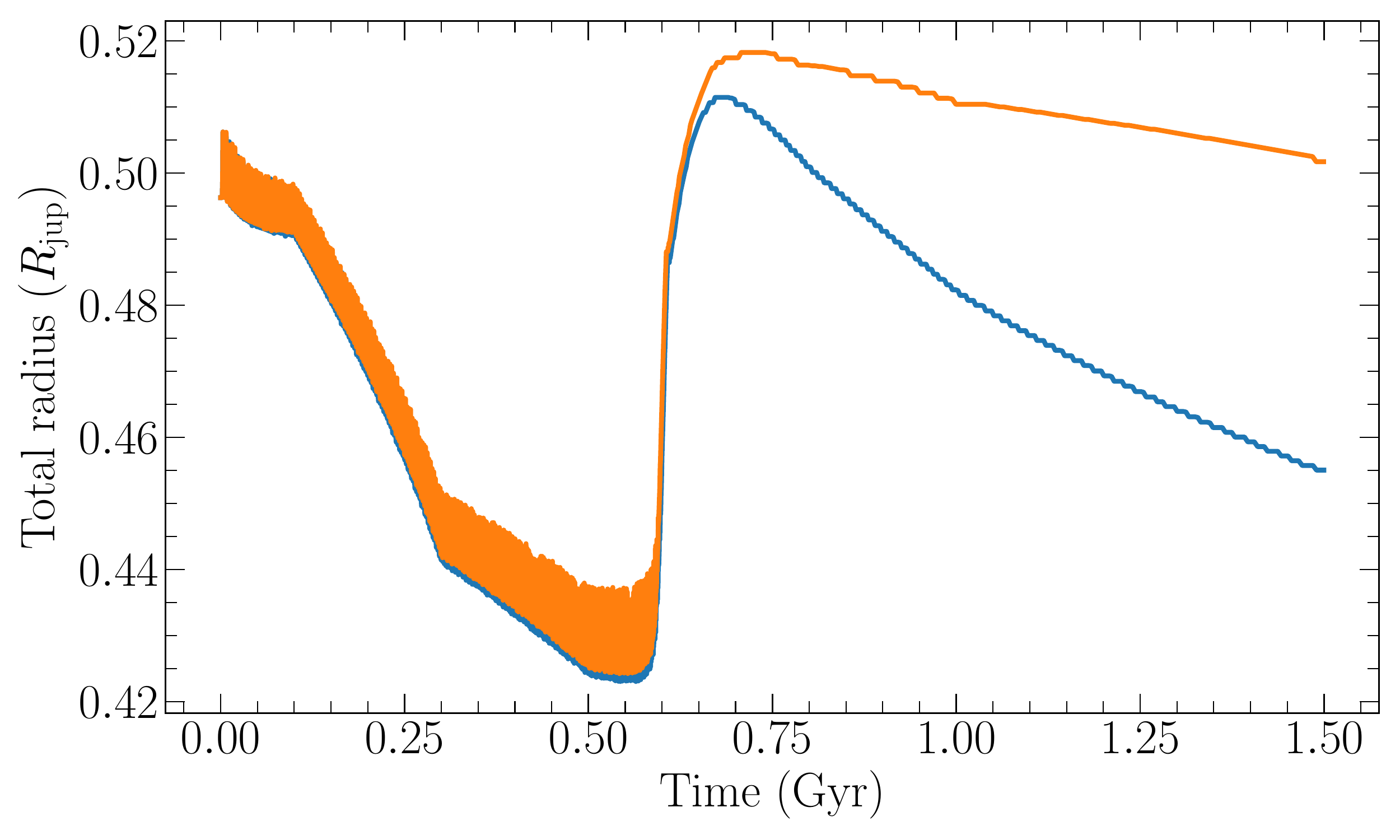}
\includegraphics[width=8.5cm]{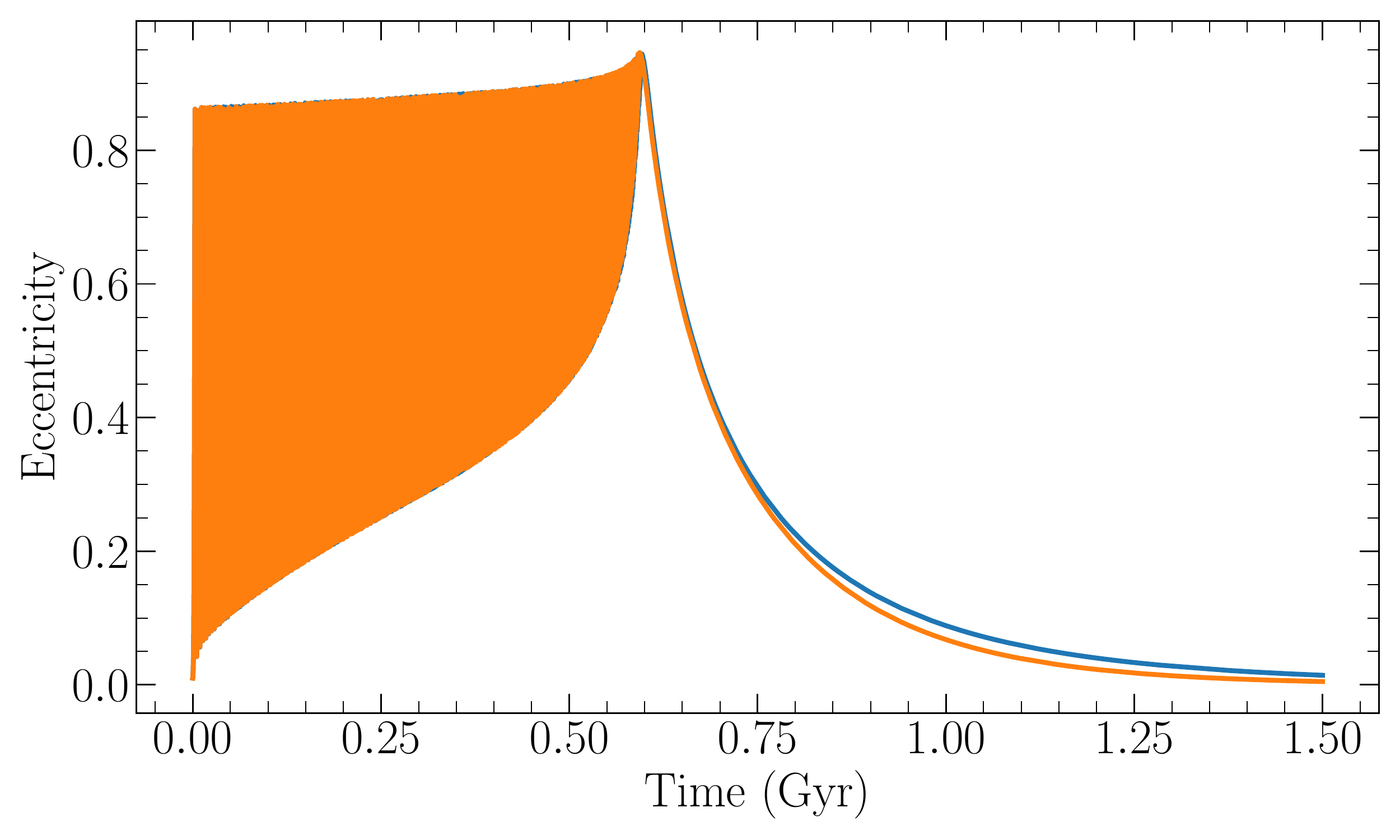}
\caption{Secular evolution of the GJ436 system. The simulation was performed including both dynamical features and an atmospheric structure. Photo-evaporation is inactive (orange). The same simulation where photo-evaporation is turned on is showed in blue for comparison. \textit{Left}: planetary radius. \textit{Right}: eccentricity of the inner orbit.}
\label{fig:GJ436_atmo_noevap}
\end{figure*}

\begin{figure}
\resizebox{\hsize}{!}{\includegraphics{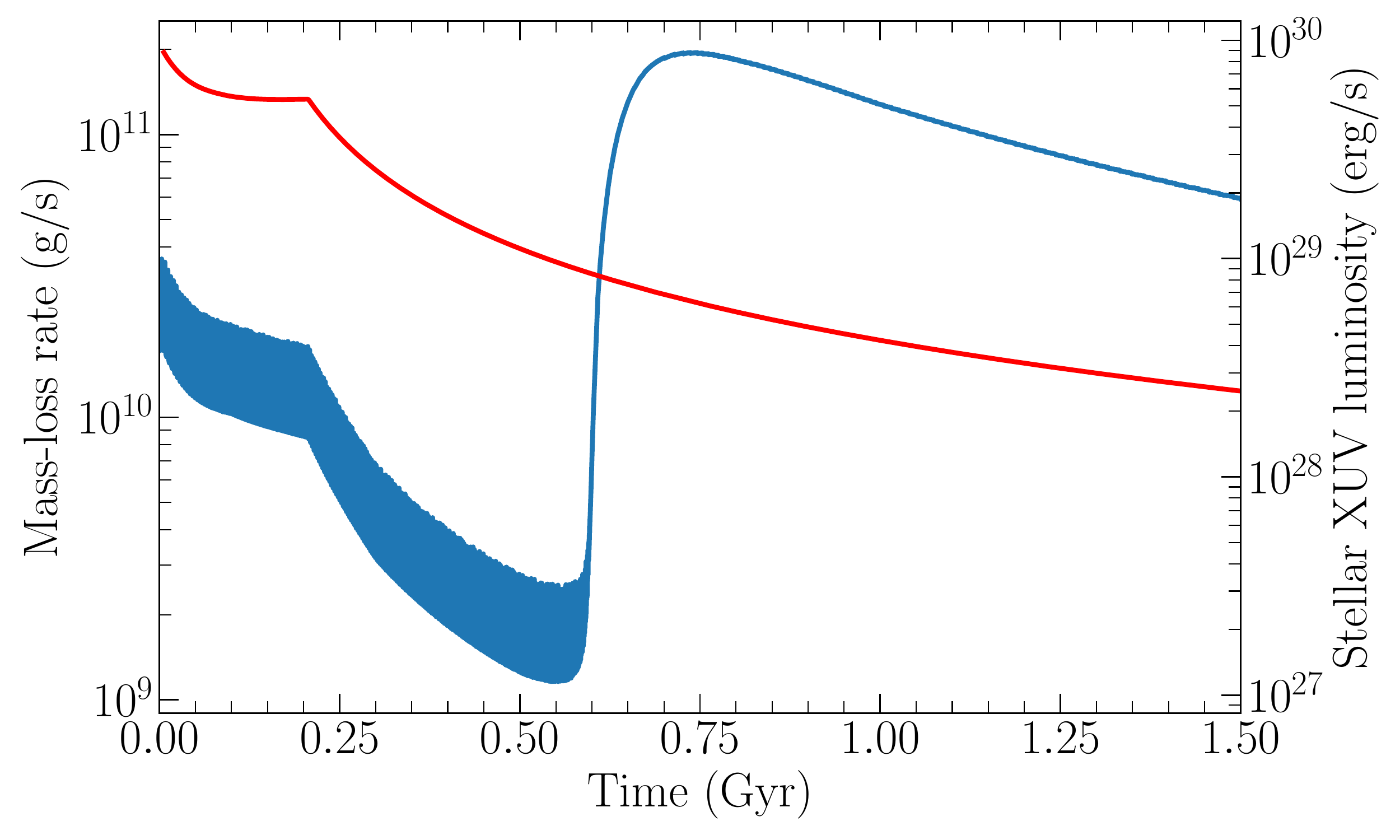}}
\caption{Evolution of the mass-loss rate of the H/He envelope of GJ436 b (blue) compared to the evolution of the XUV luminosity of GJ436 (red).}
\label{fig:GJ436_massloss}
\end{figure}

The results are presented in Fig. \ref{fig:GJ436_atmo}. The same two-part behavior of the Kozai resonance remains unaltered, but remarkably the transition occurs several billion years earlier than when considering dynamical features alone ($\tau_\mathrm{trans}\simeq$ 0.6 vs 5 Gyr). This is because the atmosphere of the planet, now taken into account in the simulation, has a strong coupling with tides. The planetary radius decreases over time due to the cooling of its rocky core steeply after the planet formation and then progressively following the secular attenuation of the stellar bolometric luminosity. On top of this global decrease, eccentricity variations due to the Kozai resonance induce high-frequency radius variations because of the secularization of the stellar bolometric flux (i.e., averaged over the inner orbit, see Eq. \ref{eq:temp}). During the high eccentricity phases, the planet spends on average more time close to the periastron. The atmosphere therefore gets much more irradiated by the star, its radius swells and pulsates with each Kozai cycle. The net effect of secularizing the bolometric flux is illustrated in Fig. \ref{fig:GJ436_atmo_sec}, which highlights the feedback of dynamical processes on the atmospheric structure. Conversely, modeling the planetary atmosphere also has an important feedback on the dynamical evolution of the system, because the planetary radius is overall larger than in the pure dynamical case. In our framework, this is due to the combination of the higher internal luminosity and the stronger intrinsic stellar luminosity during the early phases of the planetary evolution, in addition to the Kozai resonance that generates the radius pulsations. According to the simplified dynamical formalism we adopted for tides, this results in a shorter tidal-damping time scale \citep[proportional to $M_\mathrm{pl}/R_\mathrm{pl}^5$, e.g.,][]{Goldreich1966,Jackson2008}, a faster decline of the semi-major axis (Fig. \ref{fig:GJ436_atmo_sec}), and a faster thinning of the eccentricity envelope that makes the planet leave the resonance earlier. The planet radius then steeply increases after the transition, as the gaseous atmosphere suddenly heats up due to the sharp drop in semi-major axis (Fig. \ref{fig:GJ436_atmo}). Again, this change in atmospheric structure induced by the dynamical evolution feedbacks on the planet orbit, as the inflated radius accelerates the circularization compared to the pure dynamical case ($\sim$ 1 vs 3 Gyr after the transition).

As the structure of the planetary atmosphere is also affected by photo-evaporation, we ran the same simulation with this process cut-off to evaluate its net impact on the planetary evolution. Figure \ref{fig:GJ436_atmo_noevap} shows that the envelope of the radius and eccentricity variations is nearly unchanged during the first phase. The mass-loss rate shows the same oscillations as the eccentricity during the Kozai resonance (Fig. \ref{fig:GJ436_massloss}), which is a signature of the stellar XUV flux secularization (see Eq. \ref{eq:masslosssec}) and a clear feedback of the dynamical processes on photo-evaporation. However, while the mass-loss rate is large during the first hundred millions years, it steadily decreases over the first phase as a result of the attenuation of the intrinsic stellar XUV luminosity. The decrease in envelope mass ($\sim$ 1.3\%) due to evaporation during the first phase thus appears too small to compensate for the increased tidal damping due to the inflated radius. However, the increase in XUV irradiation following the sharp drop in semi-major axis after the transition pumps the atmospheric mass-loss rate back to levels comparable to the planet's early life (Fig. \ref{fig:GJ436_massloss}). Evaporation has a noticeable impact on the atmospheric structure in the second phase, decreasing the envelope mass by about 21.5\% in the Gyr following the transition. As a result, the planet radius also decreases by about 9\% more than the sole effect of the declining stellar irradiation and inner luminosity (Fig. \ref{fig:GJ436_atmo_noevap}). This is an important result, as it shows how the atmospheric evolution of close-in planets that undergo Kozai migration can still be affected by evaporation billions of years after their formation. We emphasize that the simulated present-day orbit of GJ436 b does not match its observed properties anymore, as was the case in a pure dynamical simulation starting from the chosen set of initial properties. This highlights the need to simulate the coupled atmospheric and dynamical evolution of GJ436 b, and of close-in planets in general, to get back more accurately to the original properties of their planetary systems. Such a coupled exploration will be performed for GJ436 b with the JADE code in a follow-up study.


\section{Conclusion}
\label{sect:ccl}

If the hot Neptunes desert is now a well-known feature of close-in planets, its underlying formation mechanisms remain a vast puzzle. Unveiling the mystery of its origin requires investigating a wide panel of evolutionary processes. Being designed to simulate the secular evolution of a specific planetary system over a wide range of planet and star properties, the JADE code takes us one step closer to this deep exploration.

The JADE code implements the dynamical features encoded in previous models such as tidal forces, general relativity, and the perturbing action of a distant companion \citep[e.g.,][]{Eggleton2001,Fabrycky2007,Beust2012}. The perturbing Hamiltonian is truncated up to the hexadecapole (4th order), which is, as far as we know, the highest truncation order that has been used in dynamical models. This extreme precision allows our model to properly manage complex Kozai-Lidov resonances during secular periods of time. The main novelty of the JADE code is that the dynamical evolution of the planet accounts for the presence of an atmosphere. The code coherently integrates the detailed thermodynamical structure of the atmosphere and makes it change over secular periods, as the planet dynamically evolves and is subjected to various levels of stellar irradiation. Our intention with this paper has been to rigorously describe all the equations so that the model can be easily understandable and reproducible.

We applied the JADE code to GJ436 b, representative of the warm giants forming the border of the hot Neptunes desert. During Kozai resonance, there is a strong interaction between the dynamical and atmospheric evolution of the planet, the eccentricity cycles causing the atmosphere to pulsate and conversely the inflated atmosphere strengthening tidal effects. This leads the planet to leave the resonance, migrate, and circularize its orbit several billion years earlier compared to a purely dynamical simulation. While we confirm the conclusion by \citet{Bourrier2018a} that Kozai migration can explain the eccentric and misaligned orbit of GJ436 b, their results for the original properties of the planet and its outer companion thus need to be revised considering the effect of GJ436 b's atmosphere. Furthermore, we found that photo-evaporation alters the bulk structure of GJ436 b after it gets out of the Kozai resonance, as it migrates close to the star and receives increased levels of stellar irradiation. Kozai resonance can thus delay migration and trigger strong atmospheric mass loss several billion years after a planet formation, especially as recent work suggests that the total combined X-ray and EUX stellar emission occurs mostly after the saturated phase \citep{King2021}. This changes our view of atmospheric evolution for close-in planets, which are generally thought to lose their atmosphere within the first 10 - 100 Myr after their formation and migration within the protoplanetary disk. Our study thus shows that there can be a strong coupling between the secular dynamical and atmospheric evolution of close-in exoplanets, supporting the idea that this interplay might be essential in shaping a fraction of their population. We will address this question with the JADE code by exploring the possible histories of a representative set of close-in planets on eccentric, misaligned orbits, which could have underwent delayed evaporation following Kozai migration.

Constraining the detailed simulations of the JADE code with measurements of orbital and atmospheric properties directly linked to high-eccentricity migration (e.g., the spin-orbit angle) and photo-evaporation (e.g., the mass loss rate) will allow us to disentangle further these processes. This is a complementary approach to syntheses of the whole planet population based on approximated evolutionary processes and constrained by the bulk planetary properties alone \citep[e.g.,][]{Jin2014,Owen2018}. Eventually, combining global population synthesis and highly-detailed simulations of specific systems offers the best chance to understand the roles of long-term atmospheric and dynamical evolution in shaping the different classes of close-in planets around the desert and determine their possible filiation.


\begin{acknowledgements}
We thank the referee for their appreciative report and their help in making our manuscript more to the point. We offer our thanks to Jean-Baptiste Delisle for his help with secular planetary dynamics and useful discussions about the role of the planetary atmosphere, to Nathan Hara for discussions about efficient numerical simulations, to Martin Turbet for his helpful ideas about the atmospheric integrator, to Georges King for his valuable help on the temporal evolution of the stellar fluxes,  to Rosemary Mardling for discussions about the averaged dynamical equations, and to Victor Ruelle for his help in improving the efficiency of the code. This project has received funding from the European Research Council
(ERC) under the European Union’s Horizon 2020 research and innovation
programme (project {\sc spice dune}, grant agreement No 947634; project {\sc Four
Aces}, grant agreement No 724427). V.B. acknowledges support by the Swiss National Science Foundation (SNSF) in the frame of the National Centre for Competence in Research ``PlanetS''. P.E. has received funding from the ERC under the European Union's Horizon 2020 research and innovation programme (project {\sc starex}, grant agreement No 833925).
\end{acknowledgements}


\bibliographystyle{aa}    
\bibliography{biblio}        


\begin{appendix}


\section{Orbit averaging and equations of motion derivation}

\subsection{Perturbing forces to the Keplerian equation of motion}
\label{app:fi}

The paradigm used to determine the equations of motion is a perturbative approach to the Keplerian acceleration. The explicit expressions of the forces thus involved in Eq. (\ref{eq:rdd}) are the following. First, the acceleration due to the third body reads
\begin{equation}
\label{eq:fpert}
\mathbf{f_{pert}}=GM_{\rm{pert}}\left(\frac{\mathbf{r_{pert-pl}}}{\lVert\mathbf{r_{pert-pl}}\rVert^3}-\frac{\mathbf{r_{pert-s}}}{\lVert\mathbf{r_{pert-s}}\rVert^3}\right),
\end{equation}
where $\mathbf{r_{pert-pl}}$ is the vector linking the perturber to the inner planet and $\mathbf{r_{pert-s}}$ is the vector linking the perturber to the star. $M_{\rm pert}$ is the perturber's mass. 

Then, the acceleration due to the quadrupole moment of the star, composed of its spin distortion as well as the tidal distortion produced by the inner planet, is formulated as
\begin{multline}
\label{eq:fsd}
\mathbf{f_{SD}^s}=\frac{R_{\rm{s}}^5(1+M_{\rm{pl}}/M_{\rm{s}})k_{\rm{s}}}{r^4}\Bigg\{\left[5(\mathbf{\Omega_s}\cdot\hat{\mathbf{r}})^2-\Omega_{\rm{s}}^2-\frac{6GM_{\rm{pl}}}{r^3}\right]\hat{\mathbf{r}}\\
-2(\mathbf{\Omega_s}\cdot\hat{\mathbf{r}})\mathbf{\Omega_s}\Bigg\},
\end{multline}
where $k_{\rm{s}}$ is the apsidal motion constant of the star, $\mathbf{\Omega_s}$ its spin rate, and $\Omega_{\rm{s}}$ its norm. Note that Eq. (\ref{eq:fsd}) should contain a 6 factor, instead of the 12 factor found in \citet{Mardling2002,Beust2012}. We are thankful to R.A. Mardling (private communication) for pointing this out. 

Furthermore, the acceleration produced by the tidal damping of the star also contributes to the total force:
\begin{equation}
\label{eq:ftd}
\mathbf{f_{TD}^s}=-\frac{6nk_{\rm{s}}}{Q_{\rm{s}}}\frac{M_{\rm{pl}}}{M_{\rm{s}}}\left(\frac{R_{\rm{s}}}{a}\right)^5\left(\frac{a}{r}\right)^8\times[3(\hat{\mathbf{r}}\cdot\dot{\mathbf{r}})\hat{\mathbf{r}}+(\hat{\mathbf{r}}\times\dot{\mathbf{r}}-r\mathbf{\Omega_s)\times\hat{r}}],
\end{equation}
where $a$ is the semi-major axis of the inner orbit, $n=2\pi{}/P$ the mean motion of the mutual orbit of the star and the inner planet ($P$ being the orbital period of the inner orbit), and $Q_{\rm{s}}$ the stellar tidal dissipation factor.

Similar expressions for the inner planet as in Eqs. (\ref{eq:fsd}) and (\ref{eq:ftd}): $\mathbf{f_{SD}^{pl}}$ and $\mathbf{f_{TD}^{pl}}$ have also to be taken into account and are obtained by switching the ``s'' indices by ``pl'' indices and vice-versa. The $\mathbf{f_{SD}}$ forces account for the fact that the geometry of the star and the planet are distorted by their spin and by tides. As pointed out by \citet{Eggleton1998,Mardling2002}, their secular effect is to produce an apsidal advance. The $\mathbf{f_{TD}}$ forces on the other hand generate the circularization process.

Finally, one must include the acceleration due to the post-Newtonian potential of the inner planet-star binary:
\begin{multline}
\label{eq:frel}
\mathbf{f_{rel}}=-\frac{G(M_{\rm{s}}+M_{\rm{pl}})}{r^2c^2}\Bigg\{\Bigg[(1+3\eta)\dot{\mathbf{r}}\cdot\dot{\mathbf{r}}-2(2+\eta)\frac{G(M_{\rm{s}}+M_{\rm{pl}})}{r}\\
-\frac{3}{2}\eta\dot{\mathbf{r}}^2\Bigg]\mathbf{\hat{r}}-2(2-\eta)\dot{r}\dot{\mathbf{r}}\Bigg\},
\end{multline}
where $\eta=M_{\rm{s}}M_{\rm{pl}}/(M_{\rm{s}}+M_{\rm{pl}})^2$, $c$ the speed of light, and $\dot{r}$ the norm of $\dot{\mathbf{r}}$. As for the $\mathbf{f_{SD}}$ forces, the relativistic potential produces an apsidal advance \citep{Mardling2002,Beust2012}.

We do not use Eq. (\ref{eq:rdd}) to make orbital quantities evolve over time. Instead, the previous forces are injected in Eqs. (\ref{eq:hd}) and (\ref{eq:ed}) before being orbit-averaged.

\subsection{Orbit averaging}
\label{app:avg}

In order to consider the secular system's evolution, one key step is averaging the time-evolving equations over the planetary orbits. Equations (\ref{eq:Osd}), (\ref{eq:Opd}), (\ref{eq:hd}), and (\ref{eq:ed}) depend on the planet-star distances $\mathbf{r}$ and $\mathbf{R}$ (Fig. \ref{fig:orbit}) which are very rapidly varying quantities. Averaging over the inner orbit allows us to get rid of the $\mathbf{r}$ dependency, while averaging over the outer orbit allows us to suppress the $\mathbf{R}$ dependency. To perform this averaging, a convenient change of variable is to use the ``true anomaly'' $\theta$ instead of the time $t$ as an integration variable \citep{Eggleton1998,Mardling2002}. For this purpose, it is helpful to use these parametric forms \citep{Eggleton1998}:

\begin{equation}
\label{eq:avgrv}
\mathbf{r}=r\left(\cos{\theta}\;\hat{\mathbf{e}}+\sin{\theta}\;\hat{\mathbf{q}}\right)
\end{equation}

\begin{equation}
\label{eq:avgr}
r=\frac{a(1-e^2)}{1+e\cos\theta}
\end{equation}

\begin{equation}
\dot{\mathbf{r}}=\frac{na}{\sqrt{1-e^2}}\left[-\sin\theta\;\hat{\mathbf{e}}+\left(\cos\theta+e\right)\hat{\mathbf{q}}\right]
\end{equation}

\begin{equation}
\dot{r}=\frac{nae\sin\theta}{\sqrt{1-e^2}},
\end{equation}

where $\mathbf{\hat{q}}\equiv\mathbf{\hat{h}}\times\mathbf{\hat{e}}$. These equations are particularly useful because of the convenience of the comoving $(\hat{\mathbf{e}},\hat{\mathbf{q}},\hat{\mathbf{h}})$ orthonormal frame to express the $\mathbf{h}$ and $\mathbf{e}$ equations of motion (Eqs. \ref{eq:hd} and \ref{eq:ed}), on the one hand, and because $\theta$ shows to be a good integration variable if the function contains a substantial negative power of $r$, on the other hand, which is always the case practically. One last formula is needed to perform the change of variable \citep{Eggleton1998}:

\begin{equation}
\label{eq:thetadot}
n\mathrm{d}t=\left(1-e^2\right)^{3/2}\frac{\mathrm{d}\theta}{\left(1+e\cos\theta\right)^2}.
\end{equation}

We present here an example of such an averaging. If one wants to secularize the photo-evaporative mass-loss rate (Eq. \ref{eq:massloss}) for instance, the quantity to be averaged is $1/r^2$:

\begin{align}
\label{eq:r2}
\Bigg\langle{\frac{1}{r^2}}\Bigg\rangle&=\frac{1}{P}\int_0^P\frac{\mathrm{d}t}{r^2}\notag\\
&=\int_0^{2\pi}\frac{(1+e\cos\theta)^2}{a^2(1-e^2)^2}\frac{(1-e^2)^{3/2}}{2\pi}\frac{\mathrm{d}\theta}{(1+e\cos\theta)^2}\notag\\
&=\frac{1}{a\sqrt{1-e^2}}.
\end{align}

We used here the orbital period $P=2\pi/n$ and Eqs. (\ref{eq:avgr}) and (\ref{eq:thetadot}). Equation (\ref{eq:r2}) explains then the final secularized form of the mass-loss rate in Eq. (\ref{eq:masslosssec}).

\subsection{Equations of motion secularization}
\label{app:sec}

So as to obtain the secular dynamical equations of motion, each of the forces in Eqs. (\ref{eq:fpert}) - (\ref{eq:frel}) has to be transformed into a contribution to $\dot{\mathbf{h}}$ and $\dot{\mathbf{e}}$ according to Eqs. (\ref{eq:hd}) and (\ref{eq:ed}). Afterwards, these contributions are averaged using the procedure described in Appendix \ref{app:avg}.

For the averaging of the perturber's force (Eq. \ref{eq:fpert}), a convenient way to proceed is to expand its expression in ascending powers of the ratio $a/a_{\rm{pert}}$ using Legendre polynomials \citep{Ford2000,Mardling2002}. By assuming this ratio to be small, the force is truncated to some order. Afterwards, it is averaged over the inner then the outer orbit. \citet{Eggleton2001,Fabrycky2007} did the truncation up to the second order (quadrupolar), while \citet{Ford2000,Mardling2002} used the third order (octupolar). \citet{Beust2012} recommended fourth order (hexadecapolar) to maintain high accuracy, especially on secular time scales. This is why we adopted a fourth-order truncation for the JADE code. To the best of our knowledge, this is the first time that the third and fourth order are explicitly described in the frame of secular three-body exoplanetary dynamics.

The decomposition of the force induced by the perturber is given by

\begin{equation}
\label{eq:fpertexp}
\mathbf{f_{pert}}=\frac{GM_{\rm{}pert}}{R}\sum_{l=2}^{\infty}\frac{M_{\rm{}pl}^{l-1}-\left(-M_{\rm{}s}\right)^{l-1}}{\left(M_{\rm{}s}+M_{\rm{}pl}\right)^{l-1}}\nabla_{\mathbf{r}}\left[\left(\frac{r}{R}\right)^lP_l\left(\hat{\mathbf{r}}\cdot\hat{\mathbf{R}}\right)\right],
\end{equation} 

where the $P_l$ are the ascending Legendre polynomials. Equation (\ref{eq:fpertexp}) is then truncated to 4th order in $r/R$.

It is convenient to define $x=\mathbf{r}\cdot\hat{\mathbf{R}}$ to express the different orders \citep{Mardling2002}. The quadrupole force (2nd order) is found by taking the $l=2$ contribution in Eq. (\ref{eq:fpertexp}) and using the appropriate Legendre polynomial $P_2(\Phi)=(3\Phi^2-1)/2$:

\begin{equation}
\label{eq:fpert2}
\mathbf{f_{pert,2}}=\frac{GM_{\rm{}pert}}{R^3}\left(3x\hat{\mathbf{R}}-\mathbf{r}\right).
\end{equation}

The octupole (3rd order) is found by taking the $l=3$ contribution and using the following Legendre polynomial $P_3(\Phi)=(5\Phi^3-3\Phi)/2$:

\begin{equation}
\mathbf{f_{pert,3}}=\frac{GM_{\rm{}pert}}{R^4}\frac{M_{\rm{}pl}-M_{\rm{}s}}{M_{\rm{}s}+M_{\rm{}pl}}\left[\frac{3}{2}\left(5x^2-r^2\right)\hat{\mathbf{R}}-3x\mathbf{r}\right].
\end{equation}

Finally, using the $l=4$ contribution and $P_4(\Phi)=(35\Phi^4-30\Phi^2+3)/8$ gives the hexadecapole (4th order):

\begin{equation}
\label{eq:fpert4}
\mathbf{f_{pert,4}}=\frac{GM_{\rm{}pert}}{R^5}\frac{M_{\rm{}s}^3+M_{\rm{}pl}^3}{\left(M_{\rm{}s}+M_{\rm{}pl}\right)^3}\left[\frac{5}{2}\left(7x^2-3r^2\right)x\hat{\mathbf{R}}-\frac{3}{2}\left(5x^2-r^2\right)\mathbf{r}\right].
\end{equation}

The contributions to $\dot{\mathbf{h}}$ and $\dot{\mathbf{e}}$ related to Eqs. (\ref{eq:fpert2}) - (\ref{eq:fpert4}) are averaged over the inner orbit using Eqs. (\ref{eq:avgrv}) - (\ref{eq:thetadot}). This eliminates the dependency on $\mathbf{r}$. In order to perform the averaging over the outer orbit, one can define similar parametric forms using the outer true anomaly $\theta_{\rm pert}$:

\begin{equation}
\label{eq:avgRv}
\hat{\mathbf{R}}=\cos\theta_{\rm{}pert}\;\hat{\mathbf{E}}+\sin\theta_{\rm{}pert}\;\hat{\mathbf{Q}}
\end{equation}

\begin{equation}
R=\frac{a_{\rm{}pert}(1-e_{\rm{}pert}^2)}{1+e_{\rm{}pert}\cos\theta_{\rm{}pert}}
\end{equation}

\begin{equation}
\label{eq:thetapertdot}
n_{\mathrm{pert}}\mathrm{d}t=\left(1-e_{\mathrm{pert}}^2\right)^{3/2}\frac{\mathrm{d}\theta_{\mathrm{pert}}}{\left(1+e_{\mathrm{pert}}\cos\theta_{\mathrm{pert}}\right)^2}.
\end{equation}

We define here a direct orthonormal frame $(\hat{\mathbf{E}},\hat{\mathbf{Q}},\hat{\mathbf{H}})$ for the outer orbit as well. The projection coefficients between the two frames are $\hat{E}_x\equiv\hat{\mathbf{E}}\cdot\hat{\mathbf{x}}$, $\hat{Q}_x\equiv\hat{\mathbf{Q}}\cdot\hat{\mathbf{x}}$, and $\hat{H}_x\equiv\hat{\mathbf{H}}\cdot\hat{\mathbf{x}}$, where $\hat{\mathbf{x}}$ is either $\hat{\mathbf{e}}$, $\hat{\mathbf{q}}$, or $\hat{\mathbf{h}}$. With Eqs. (\ref{eq:avgRv}) - (\ref{eq:thetapertdot}), one can average the 3 contributions to the perturber's force over the outer orbit to remove the dependency on $\mathbf{\hat{R}}$ and $R$. We note that we derived a useful formula which allows very welcome simplifications in the calculation of the averaged perturber's force. It is related to the orthonormality of the two bases $(\hat{\mathbf{E}},\hat{\mathbf{Q}},\hat{\mathbf{H}})$ and $(\hat{\mathbf{e}},\hat{\mathbf{q}},\hat{\mathbf{h}})$. Its expression is the following:

\begin{equation}
\hat{E}_x\hat{E}_y+\hat{Q}_x\hat{Q}_y+\hat{H}_x\hat{H}_y=\delta_{xy},
\end{equation}

where $x$ and $y$ are either e, q, or h, and $\delta_{xy}$ is the Kronecker delta.

Tidal and relativistic terms (Eqs. \ref{eq:fsd} - \ref{eq:frel}) only have to be averaged over the inner orbit since they do not depend on the outer orbit. \citet{Mardling2002} gave explicit expressions for the averaged equations, which are consistent with our results.

At this level of complexity, numerical computing is required to derived the many averaged equations, which is why we made use of \textsc{Mathematica} \citep{Mathematica}. The final secularized equations of motion can be written as

\begin{align}
\label{eq:hdavg}
\left\langle\left\langle{\frac{\mathrm{d}\mathbf{h}}{\mathrm{d}t}}\right\rangle\right\rangle=&\left\langle\left\langle{\mathbf{d_{pert,2}}}\right\rangle\right\rangle+\left\langle\left\langle{\mathbf{d_{pert,3}}}\right\rangle\right\rangle+\left\langle\left\langle{\mathbf{d_{pert,4}}}\right\rangle\right\rangle \\
&+\left\langle{\mathbf{d_{SD}^s}}\right\rangle+\left\langle{\mathbf{d_{SD}^{pl}}}\right\rangle+\left\langle{\mathbf{d_{TD}^s}}\right\rangle+\left\langle{\mathbf{d_{TD}^{pl}}}\right\rangle \nonumber
\end{align}
\begin{align}
\left\langle\left\langle{\frac{\mathrm{d}\mathbf{e}}{\mathrm{d}t}}\right\rangle\right\rangle=&\left\langle\left\langle{\mathbf{g_{pert,2}}}\right\rangle\right\rangle+\left\langle\left\langle{\mathbf{g_{pert,3}}}\right\rangle\right\rangle+\left\langle\left\langle{\mathbf{g_{pert,4}}}\right\rangle\right\rangle \\
&+\left\langle{\mathbf{g_{SD}^s}}\right\rangle+\left\langle{\mathbf{g_{SD}^{pl}}}\right\rangle+\left\langle{\mathbf{g_{TD}^s}}\right\rangle+\left\langle{\mathbf{g_{TD}^{pl}}}\right\rangle+\Big\langle{\mathbf{g_{rel}}}\Big\rangle \nonumber
\end{align}
\begin{equation}
I_{\rm{s}}\;\Bigg\langle\Bigg\langle{\frac{\mathrm{d}\mathbf{\Omega_s}}{\mathrm{d}t}}\Bigg\rangle\Bigg\rangle=-\frac{M_{\rm{s}}M_{\rm{pl}}}{M_{\rm{s}}+M_{\rm{pl}}}\Big(\Big\langle{\mathbf{d_{SD}^{s}}}\Big\rangle+\Big\langle{\mathbf{d_{TD}^{s}}}\Big\rangle\Big)
\end{equation}
\begin{equation}
\label{eq:Opdavg}
I_{\rm{pl}}\;\Bigg\langle\Bigg\langle{\frac{\mathrm{d}\mathbf{\Omega_{pl}}}{\mathrm{d}t}}\Bigg\rangle\Bigg\rangle=-\frac{M_{\rm{s}}M_{\rm{pl}}}{M_{\rm{s}}+M_{\rm{pl}}}\Big(\Big\langle{\mathbf{d_{SD}^{pl}}}\Big\rangle+\Big\langle{\mathbf{d_{TD}^{pl}}}\Big\rangle\Big).
\end{equation}

The different contributions are as follows. First, the quadrupolar contribution of the force due to the third body reads

\begin{align}
\label{eq:dquad}
\Big\langle\Big\langle{\mathbf{d_{pert,2}}}\Big\rangle\Big\rangle=&\frac{3GM_{\rm{}pert}a^2}{4a_{\rm{}pert}^3(1-e_{\rm{}pert}^2)^{3/2}}\;\Bigg\{-(1-e^2)\hat{H}_{\rm{}q}\hat{H}_{\rm{}h}\hat{\mathbf{e}} \\
&+(1+4e^2)\hat{H}_{\rm{}e}\hat{H}_{\rm{}h}\hat{\mathbf{q}}-5e^2\hat{H}_{\rm{}e}\hat{H}_{\rm{}q}\hat{\mathbf{h}}\Bigg\} \nonumber
\end{align}

\begin{align}
\label{eq:gquad}
\Big\langle\Big\langle{\mathbf{g_{pert,2}}}\Big\rangle\Big\rangle=&\frac{3nM_{\rm{}pert}a^3e\sqrt{1-e^2}}{4(M_{\rm{}s}+M_{\rm{}pl})a_{\rm{}pert}^3(1-e_{\rm{}pert}^2)^{3/2}}\;\Bigg\{5\hat{H}_{\rm{}e}\hat{H}_{\rm{}q}\hat{\mathbf{e}} \\
&+(-3+5\hat{H}_{\rm{}q}^2+4\hat{H}_{\rm{}h}^2)\hat{\mathbf{q}}+\hat{H}_{\rm{}q}\hat{H}_{\rm{}h}\hat{\mathbf{h}}\Bigg\}. \nonumber
\end{align}

The octupolar contribution of the force due to the third body is

\begin{align}
\Big\langle\Big\langle{\mathbf{d_{pert,3}}}\Big\rangle\Big\rangle=&\frac{15GM_{\rm{}pert}(M_{\rm{}s}-M_{\rm{}pl})a^3ee_{\rm{}pert}}{64(M_{\rm{}s}+M_{\rm{}pl})a_{\rm{}pert}^4(1-e_{\rm{}pert}^2)^{5/2}} \\
&\Bigg\{\alpha_{\rm{}e}\mathbf{\hat{e}}+\alpha_{\rm{}q}\mathbf{\hat{q}}+\alpha_{\rm{}h}\mathbf{\hat{h}}\Bigg\} \nonumber
\end{align}

\begin{align}
\Big\langle\Big\langle{\mathbf{g_{pert,3}}}\Big\rangle\Big\rangle=&\frac{15nM_{\rm{}pert}(M_{\rm{}s}-M_{\rm{}pl})a^4\sqrt{1-e^2}e_{\rm{}pert}}{64(M_{\rm{}s}+M_{\rm{}pl})^2a_{\rm{}pert}^4(1-e_{\rm{}pert}^2)^{5/2}} \\
&\Bigg\{\beta_{\rm{}e}\mathbf{\hat{e}}+\beta_{\rm{}q}\mathbf{\hat{q}}+\beta_{\rm{}h}\mathbf{\hat{h}}\Bigg\}. \nonumber
\end{align}

We only present here the prefactors to maintain readability. The coefficients $\alpha_x$ and $\beta_x$ are expressed explicitly in Table \ref{tab:pert}.

The hexadecapolar contribution of the force due to the third body is

\begin{align}
\Big\langle\Big\langle{\mathbf{d_{pert,4}}}\Big\rangle\Big\rangle=&\frac{45GM_{\rm{}pert}(M_{\rm{}s}^3+M_{\rm{}pl}^3)a^4}{256(M_{\rm{}s}+M_{\rm{}pl})^3a_{\rm{}pert}^5(1-e_{\rm{}pert}^2)^{7/2}}\\
&\Bigg\{\gamma_{\rm{}e}\mathbf{\hat{e}}+\gamma_{\rm{}q}\mathbf{\hat{q}}+\gamma_{\rm{}h}\mathbf{\hat{h}}\Bigg\} \nonumber
\end{align}

\begin{align}
\label{eq:ghex}
\Big\langle\Big\langle{\mathbf{g_{pert,4}}}\Big\rangle\Big\rangle=&\frac{45nM_{\rm{}pert}(M_{\rm{}s}^3+M_{\rm{}pl}^3)a^5e\sqrt{1-e^2}}{256(M_{\rm{}s}+M_{\rm{}pl})^4a_{\rm{}pert}^5(1-e_{\rm{}pert}^2)^{7/2}}\\
&\Bigg\{\delta_{\rm{}e}\mathbf{\hat{e}}+\delta_{\rm{}q}\mathbf{\hat{q}}+\delta_{\rm{}h}\mathbf{\hat{h}}\Bigg\}. \nonumber
\end{align}

The coefficients $\gamma_x$ and $\delta_x$ can be found in Table \ref{tab:pert} as well.

Then, the averaging of the spin distortion of the star can be expressed as

\begin{equation}
\label{eq:dsd}
\Big\langle{\mathbf{d_{SD}^s}}\Big\rangle=\frac{k_{\rm{}s}(1+M_{\rm{}pl}/M_{\rm{}s})R_{\rm{}s}^5}{a^3(1-e^2)^{3/2}}\;\Bigg\{-\Omega_{\rm{}s,q}\Omega_{\rm{}s,h}\hat{\mathbf{e}}+\Omega_{\rm{}s,e}\Omega_{\rm{}s,h}\hat{\mathbf{q}}\;\Bigg\}
\end{equation}

\begin{align}
\label{eq:gsd}
\Big\langle{\mathbf{g_{SD}^s}}\Big\rangle=&\frac{nk_{\rm{}s}R_{\rm{}s}^5e}{GM_{\rm{}s}a^2(1-e^2)^2}\;\Bigg\{\Big[\frac{15}{8}\frac{8+12e^2+e^4}{(1-e^2)^3}\frac{GM_{\rm{}pl}}{a^3} \\
&-\frac{1}{2}(\Omega_{\rm{}s,e}^2+\Omega_{\rm{}s,q}^2-2\Omega_{\rm{}s,h}^2)\Big]\hat{\mathbf{q}}+\Omega_{\rm{}s,q}\Omega_{\rm{}s,h}\hat{\mathbf{h}}\Bigg\}. \nonumber
\end{align}

Here, $\Omega_{\mathrm{s,}x}\equiv\mathbf{\Omega_{\rm{}s}}\cdot\hat{\mathbf{x}}$ where $\hat{\mathbf{x}}$ is either $\hat{\mathbf{e}}$, $\hat{\mathbf{q}}$, or $\hat{\mathbf{h}}$.

Furthermore, the tidal damping of the star contribution is now formulated as

\begin{align}
\Big\langle{\mathbf{d_{TD}^s}}\Big\rangle=&\frac{3nk_{\rm{}s}M_{\rm{}pl}R_{\rm{}s}^5}{8Q_{\rm{}s}M_{\rm{}s}a^3(1-e^2)^{9/2}}\;\Bigg\{(8+12e^2+e^4)\Omega_{\rm{}s,e}\hat{\mathbf{e}} \\
&+(8+36e^2+5e^4)\Omega_{\rm{}s,q}\hat{\mathbf{q}}-\Big[\frac{16+5e^2(24+18e^2+e^4)}{(1-e^2)^{3/2}}n \nonumber \\
&-2(8+3e^2(8+e^2))\Omega_{\rm{}s,h}\Big]\hat{\mathbf{h}}\Bigg\} \nonumber
\end{align}

\begin{align}
\label{eq:gtf}
\Big\langle{\mathbf{g_{TD}^s}}\Big\rangle=&\frac{3n^2k_{\rm{}s}R_{\rm{}s}^5e}{8Q_{\rm{}s}GM_{\rm{}s}(1+M_{\rm{}s}/M_{\rm{}pl})a^2(1-e^2)^5} \\
&\Bigg\{\Big[-\frac{9}{4}\frac{64+5e^2(48+24e^2+e^4)}{(1-e^2)^{3/2}}n \nonumber \\
&+11(8+12e^2+e^4)\Omega_{\rm{}s,h}\Big]\hat{\mathbf{e}}-(8+12e^2+e^4)\Omega_{\rm{}s,e}\hat{\mathbf{h}}\Bigg\}. \nonumber
\end{align}

Similar expressions for the main planet as in Eqs. (\ref{eq:dsd}) - (\ref{eq:gtf}) also have to be taken into account and are obtained by switching the ``s'' indices by ``pl'' indices and vice-versa.

Finally, the averaged post-Newtonian contribution is

\begin{equation}
\label{eq:grel}
\Big\langle{\mathbf{g_{rel}}}\Big\rangle=\frac{3a^2n^3e}{c^2(1-e^2)}\hat{\mathbf{q}}.
\end{equation}

\begin{table*}
\caption{Coefficients involved in the expression of the octupolar and hexadecapolar contributions of the perturber's force.}
\label{tab:pert}
\centering
\begin{tabular}{c c}
\hline \hline \\
Coefficient & Expression \\
& \\
\hline \\

$\alpha_{\rm e}$ & $10\left(1-e^2\right)\left[\hat{Q}_{\rm{}e}\left(\hat{E}_{\rm{}h}\hat{Q}_{\rm{}q}+\hat{E}_{\rm{}q}\hat{Q}_{\rm{}h}\right)+\hat{E}_{\rm{}e}\left(3\hat{E}_{\rm{}q}\hat{E}_{\rm{}h}+\hat{Q}_{\rm{}q}\hat{Q}_{\rm{}h}\right)\right]$ \\

& \\

$\alpha_{\rm q}$ & $-\left[\hat{E}_{\rm{}h}\left(-1+8e^2+\left(45+60e^2\right)\hat{H}_{\rm{}h}^2-5\left(2+5e^2\right)\hat{Q}_{\rm{}q}^2\right)+5\left(2+5e^2\right)\hat{E}_{\rm{}q}\left(3\hat{H}_{\rm{}q}\hat{H}_{\rm{}h}+\hat{Q}_{\rm{}q}\hat{Q}_{\rm{}h}\right)\right]$ \\

& \\

$\alpha_{\rm h}$ & $\left[-10\left(1+6e^2\right)\hat{E}_{\rm{}h}\hat{Q}_{\rm{}q}\hat{Q}_{\rm{}h}+\hat{E}_{\rm{}q}\left(-11+15\hat{H}_{\rm{}h}^2+10\hat{Q}_{\rm{}h}^2+3e^2\left(-29+35\hat{H}_{\rm{}q}^2+30\hat{H}_{\rm{}h}^2+20\hat{Q}_{\rm{}h}^2\right)\right)\right]$ \\

& \\
\hline \\

$\beta_{\rm e}$ & $\left[10\left(1+6e^2\right)\hat{E}_{\rm{}h}\hat{Q}_{\rm{}q}\hat{Q}_{\rm{}h}-\hat{E}_{\rm{}q}\left(-11+15\hat{H}_{\rm{}h}^2+10\hat{Q}_{\rm{}h}^2+3e^2\left(-29+35\hat{H}_{\rm{}q}^2+30\hat{H}_{\rm{}h}^2+20\hat{Q}_{\rm{}h}^2\right)\right)\right]$ \\

& \\

$\beta_{\rm q}$ & $\left[\left(-10+30e^2\right)\hat{E}_{\rm{}h}\hat{Q}_{\rm{}e}\hat{Q}_{\rm{}h}+\hat{E}_{\rm{}e}\left(-11+15\hat{H}_{\rm{}h}^2+10\hat{Q}_{\rm{}h}^2+3e^2\left(-17+35\hat{H}_{\rm{}q}^2+20\hat{H}_{\rm{}h}^2-10\hat{Q}_{\rm{}h}^2\right)\right)\right]$ \\

& \\

$\beta_{\rm h}$ & $10e^2\left[\hat{Q}_{\rm{}e}\left(\hat{E}_{\rm{}h}\hat{Q}_{\rm{}q}+\hat{E}_{\rm{}q}\hat{Q}_{\rm{}h}\right)+\hat{E}_{\rm{}e}\left(3\hat{E}_{\rm{}q}\hat{E}_{\rm{}h}+\hat{Q}_{\rm{}q}\hat{Q}_{\rm{}h}\right)\right]$ \\

& \\
\hline \\

$\gamma_{\rm e}$ & $\Big(1-e^2\Big)\Big[98e^2e_{\rm{}pert}^2\hat{H}_{\rm{}q}^2\hat{Q}_{\rm{}q}\hat{Q}_{\rm{}h}+\hat{H}_{\rm{}h}\hat{H}_{\rm{}q}\Big(e^2\Big(e_{\rm{}pert}^2\Big(210\hat{H}_{\rm{}h}^2+98\hat{Q}_{\rm{}q}^2+84\hat{Q}_{\rm{}h}^2-221\Big)+84\hat{H}_{\rm{}h}^2-78\Big)$ \\
& $+e_{\rm{}pert}^2\Big(35\hat{H}_{\rm{}h}^2+14\hat{Q}_{\rm{}h}^2-17\Big)+14\hat{H}_{\rm{}h}^2-6\Big)+2e_{\rm{}pert}^2\hat{Q}_{\rm{}q}\hat{Q}_{\rm{}h}\Big(\Big(42e^2+7\Big)\hat{H}_{\rm{}h}^2-13e^2-1\Big)+49e^2\Big(5e_{\rm{}pert}^2+2\Big)\hat{H}_{\rm{}h}\hat{H}_{\rm{}q}^3\Big]$ \\

& \\

$\gamma_{\rm q}$ & $-\Big[\hat{H}_{\rm{}e}\Big(98e^2\Big(2e^2+1\Big)e_{\rm{}pert}^2\hat{H}_{\rm{}h}\hat{Q}_{\rm{}q}^2+98e^2\Big(2e^2+1\Big)e_{\rm{}pert}^2\hat{H}_{\rm{}q}\hat{Q}_{\rm{}q}\hat{Q}_{\rm{}h}+49e^2\Big(2e^2+1\Big)\Big(5e_{\rm{}pert}^2+2\Big)\hat{H}_{\rm{}q}^2\hat{H}_{\rm{}h}$ \\
& $+14\Big(22e^4+19e^2+1\Big)e_{\rm{}pert}^2\hat{H}_{\rm{}h}\hat{Q}_{\rm{}h}^2+7\Big(8e^4+12e^2+1\Big)\Big(5e_{\rm{}pert}^2+2\Big)\hat{H}_{\rm{}h}^3-\hat{H}_{\rm{}h}\Big(76e^4+86e^2$ \\
& $+\Big(346e^4+309e^2+17\Big)e_{\rm{}pert}^2+6\Big)\Big)-2e_{\rm{}pert}^2\hat{Q}_{\rm{}e}\hat{Q}_{\rm{}h}\Big(-20e^4-2e^2+7\Big(6e^4-5e^2-1\Big)\hat{H}_{\rm{}h}^2+1\Big)\Big]$ \\

& \\

$\gamma_{\rm h}$ & $-7e^2\Big[2e_{\rm{}pert}^2\hat{Q}_{\rm{}e}\Big(\Big(7-7e^2\Big)\hat{H}_{\rm{}q}\hat{H}_{\rm{}h}\hat{Q}_{\rm{}h}+\Big(e^2+2\Big)\hat{Q}_{\rm{}q}\Big)+\hat{H}_{\rm{}e}\Big(\hat{H}_{\rm{}q}\Big(e^2\Big(e_{\rm{}pert}^2\Big(70\hat{H}_{\rm{}h}^2+84\hat{Q}_{\rm{}q}^2+42\hat{Q}_{\rm{}h}^2-95\Big)+28\hat{H}_{\rm{}h}^2-22\Big)$ \\
& $+7\Big(5e_{\rm{}pert}^2+2\Big)\hat{H}_{\rm{}h}^2-e_{\rm{}pert}^2-2\Big)+14\Big(2e^2+1\Big)e_{\rm{}pert}^2\hat{H}_{\rm{}h}\hat{Q}_{\rm{}q}\hat{Q}_{\rm{}h}+21e^2\Big(5e_{\rm{}pert}^2+2\Big)\hat{H}_{\rm{}q}^3\Big)\Big]$ \\

& \\
\hline \\

$\delta_{\rm e}$ & $7\Big[2e_{\rm{}pert}^2\hat{Q}_{\rm{}e}\Big(\Big(7-7e^2\Big)\hat{H}_{\rm{}q}\hat{H}_{\rm{}h}\hat{Q}_{\rm{}h}+\Big(e^2+2\Big)\hat{Q}_{\rm{}q}\Big)+\hat{H}_{\rm{}e}\Big(\hat{H}_{\rm{}q}\Big(e^2\Big(e_{\rm{}pert}^2\Big(70\hat{H}_{\rm{}h}^2+84\hat{Q}_{\rm{}q}^2+42\hat{Q}_{\rm{}h}^2-95\Big)+28\hat{H}_{\rm{}h}^2-22\Big)$ \\
& $+7\Big(5e_{\rm{}pert}^2+2\Big)\hat{H}_{\rm{}h}^2-e_{\rm{}pert}^2-2\Big)+14\Big(2e^2+1\Big)e_{\rm{}pert}^2\hat{H}_{\rm{}h}\hat{Q}_{\rm{}q}\hat{Q}_{\rm{}h}+21e^2\Big(5e_{\rm{}pert}^2+2\Big)\hat{H}_{\rm{}q}^3\Big)\Big]$ \\

& \\

$\delta_{\rm q}$ & $-\Big[588e^2e_{\rm{}pert}^2\hat{H}_{\rm{}q}^2\hat{Q}_{\rm{}q}^2+196\Big(4e^2+1\Big)e_{\rm{}pert}^2\hat{H}_{\rm{}h}\hat{H}_{\rm{}q}\hat{Q}_{\rm{}q}\hat{Q}_{\rm{}h}+56\Big(4e^2+3\Big)e_{\rm{}pert}^2\hat{H}_{\rm{}h}^2\hat{Q}_{\rm{}h}^2+147e^2\Big(5e_{\rm{}pert}^2+2\Big)\hat{H}_{\rm{}q}^4$ \\
& $+49\Big(4e^2+1\Big)\Big(5e_{\rm{}pert}^2+2\Big)\hat{H}_{\rm{}h}^2\hat{H}_{\rm{}q}^2-7\hat{H}_{\rm{}q}^2\Big(2e^2\Big(53e_{\rm{}pert}^2+22\Big)+e_{\rm{}pert}^2+2\Big)+14\Big(4e^2+3\Big)\Big(5e_{\rm{}pert}^2+2\Big)\hat{H}_{\rm{}h}^4$ \\
& $-\hat{H}_{\rm{}h}^2\Big(4e^2\Big(75e_{\rm{}pert}^2+38\Big)+211e_{\rm{}pert}^2+86\Big)+28\Big(e^2+1\Big)e_{\rm{}pert}^2\hat{Q}_{\rm{}q}^2+4\Big(20e^2+1\Big)e_{\rm{}pert}^2\hat{Q}_{\rm{}h}^2+15e^2e_{\rm{}pert}^2+46e^2-e_{\rm{}pert}^2+10\Big]$ \\

& \\

$\delta_{\rm h}$ & $-\Big[98e^2e_{\rm{}pert}^2\hat{H}_{\rm{}q}^2\hat{Q}_{\rm{}q}\hat{Q}_{\rm{}h}+\hat{H}_{\rm{}h}\hat{H}_{\rm{}q}\Big(e^2\Big(e_{\rm{}pert}^2\Big(210\hat{H}_{\rm{}h}^2+98\hat{Q}_{\rm{}q}^2+84\hat{Q}_{\rm{}h}^2-221\Big)+84\hat{H}_{\rm{}h}^2-78\Big)+e_{\rm{}pert}^2\Big(35\hat{H}_{\rm{}h}^2+14\hat{Q}_{\rm{}h}^2-17\Big)$ \\
& $+14\hat{H}_{\rm{}h}^2-6\Big)+2e_{\rm{}pert}^2\hat{Q}_{\rm{}q}\hat{Q}_{\rm{}h}\Big(\Big(42e^2+7\Big)\hat{H}_{\rm{}h}^2-13e^2-1\Big)+49e^2\Big(5e_{\rm{}pert}^2+2\Big)\hat{H}_{\rm{}h}\hat{H}_{\rm{}q}^3\Big]$ \\

& \\
\hline
\end{tabular}
\end{table*}


\section{Numerical integration}
\label{app:numerical}

Two distinct numerical integrators are implemented in the JADE code. The first one temporally solves the secular equations of motion, while the second one spatially integrates the atmospheric structure in order to determine the planetary radius.

Equations (\ref{eq:hdavg}) - (\ref{eq:Opdavg}) are time-integrated by the JADE code on secular time scales using a fifth-order Runge-Kutta scheme with an adaptive time step. The equations of motion integrator relies on the Dormand-Prince method and takes advantage of a ``just-in-time'' compilation and a C-based execution to greatly speed up computation time. In this way, a secular dynamical integration over 10 Gyr does not last more than a few hours. The stability of the integrator is first guaranteed by the dimensionless units used in the code. This makes all quantities close to unity and ensures that the time step control is consistent with the evolution of all the parameters at once. In addition, the relative and absolute tolerance of the integrator are respectively set to $10^{-6}$ and $10^{-12}$, which ensures a small-enough time step to correctly match the complex involved dynamics. By doing so, Kozai oscillations do not change if a stricter tolerance is imposed, and angular momentum relative errors are satisfyingly always lower than $\sim 10^{-6}$.

The integration of the 1D atmospheric equations is also performed using a fifth-order Runge-Kutta scheme with an adaptive step. Dimensionless units are again adopted and the same tolerances as for the dynamical integrator are consistently used. For each radius determination procedure, a parallelized grid of atmospheric integrations is performed. The JADE code then retains the planetary radius which yields a mass at the center sufficiently close to zero: $M(r=0)<10^{-3} \times M_{\rm core}$. This threshold value has been chosen such that the derived radius no longer varies if a more constraining value is imposed. This determination of $R_{\rm pl}$ is computationally expensive, and it is thus evaluated with a time step that accounts for the main processes governing its evolution:

\begin{equation}
\tau_{\rm{eval}}=\eta\min\left(\tau_\mathrm{evap},\tau_\mathrm{Kozai}\right),
\end{equation}

where $\tau_{\rm evap}=\vert M_{\rm env}/\dot{M}_{\rm env} \vert$ is the characteristic photo-evaporation time scale, $\tau_{\rm Kozai}$ is the characteristic time scale of Kozai oscillations (Eq. \ref{eq:taukozai}) that induce radius pulsations, and $\eta$ is a safety factor (typically set to $10^{-2}$). We note that $\tau_{\rm eval}$ is re-evaluated every time $R_{\rm pl}$ is calculated. One can see in Figs. \ref{fig:atmoevap} and \ref{fig:GJ436_atmo} that this approach allows the evolution of the envelope to be smoothly determined.


\section{Hydrogen/helium equations of state}
\label{app:saumon}

\citet{Saumon1995} provided useful equations of state (EOS) for hydrogen and helium. They appear in the form of tables of values where each row gives the values of thermodynamical parameters in a physical configuration. Hence, given the values of the two independent parameters $T$ and $P$, one can interpolate the value of a thermodynamical parameter such as $\rho$ or $\nabla_{\rm conv}$. However, the EOS are given as two independent tables (one for hydrogen and one for helium). If we want to mix both, we have to correctly combine the derived values for hydrogen alone and for helium alone using the mass fraction of helium $Y_{\rm He}$.

The first thermodynamical quantity interpolated by the JADE code is the density $\rho$. The density can be seen as an inverse specific volume, which is an extensive quantity. Thus, one has to combine the interpolated value for hydrogen alone $\rho^{\rm{H}}$ and for helium alone $\rho^{\rm{He}}$ in the following way \citep{Saumon1995}:

\begin{equation}
\frac{1}{\rho(P, T)}=\frac{1-Y_{\mathrm{He}}}{\rho^{\mathrm{H}}(P, T)}+\frac{Y_{\mathrm{He}}}{\rho^{\mathrm{He}}(P, T)}.
\end{equation}

The second thermodynamical quantity interpolated by the JADE code is the convective gradient $\nabla_{\rm conv}$. However, its value for a H/He mix cannot be calculated using the additive-volume rule. It has to be calculated from its definition:

\begin{equation}
\nabla_{\mathrm{conv}}=\left.\frac{\partial \log T}{\partial \log P}\right|_{S}=-\frac{S_{P}}{S_{T}}.
\end{equation}

Here, $S_T$ and $S_P$ are respectively the partial derivatives of the entropy's logarithm $\log S$ with respect to $\log T$ and $\log P$. Their mixed expression can be calculated from the following expressions \citep{Saumon1995}:

\begin{equation}
S_{T}=(1-Y_{\mathrm{He}}) \frac{S}{S^{\mathrm{H}}} S_{T}^{\mathrm{H}}+Y_{\mathrm{He}} \frac{S}{S^{\mathrm{He}}} S_{T}^{\mathrm{He}}+\left.\frac{S_{\mathrm{mix}}}{S} \frac{\partial \log S_{\mathrm{mix}}}{\partial \log T}\right|_{P}
\end{equation}

\begin{equation}
S_{P}=(1-Y_{\mathrm{He}}) \frac{S}{S^{\mathrm{H}}} S_{P}^{\mathrm{H}}+Y_{\mathrm{He}} \frac{S}{S^{\mathrm{He}}} S_{P}^{\mathrm{He}}+\left.\frac{S_{\mathrm{mix}}}{S} \frac{\partial \log S_{\mathrm{mix}}}{\partial \log P}\right|_{T}.
\end{equation}

The total entropy is calculated using the additive rule, as it is an extensive quantity: $S=(1-Y_{\rm He})S^{\rm H}+Y_{\rm He}S^{\rm He}$. Finally, the mixing entropy $S_{\rm mix}$ can be calculated as follows \citep{Saumon1995}:

\begin{multline} 
\frac{S_{\mathrm{mix}}}{k_{\mathrm{B}}}=\frac{1-Y_{\mathrm{He}}}{m_{\mathrm{H}}} \frac{2}{\left(1+X_{\mathrm{H}}+3 X_{\mathrm{H}_{2}}\right)}\Bigg\{\ln (1+\beta \gamma)-X_{e}^{\mathrm{H}} \ln (1+\delta)\\
+\beta \gamma\left[\ln (1+1 / \beta \gamma)-X_{e}^{\mathrm{He}} \ln (1+1 / \delta)\right]\Bigg\}.
\end{multline}

$k_{\rm B}$ is the Boltzmann constant. The concentrations $X_i$ are read from the tables. The $\beta$, $\gamma$ and $\delta$ coefficients are given by \citep{Saumon1995,Broeg2009}:

\begin{equation} 
\beta=\frac{m_{\mathrm{H}}}{m_{\mathrm{He}}} \frac{Y_{\mathrm{He}}}{1-Y_{\mathrm{He}}}
\end{equation}

\begin{equation}
\gamma=\frac{3}{2} \frac{\left(1+X_{\mathrm{H}}+3 X_{\mathrm{H}_{2}}\right)}{\left(1+2 X_{\mathrm{He}}+X_{\mathrm{He}^{+}}\right)} 
\end{equation}

\begin{equation}
\delta=\frac{2}{3}\frac{\left(2-2 X_{\mathrm{He}}-X_{\mathrm{He}^{+}}\right)}{\left(1-X_{\mathrm{H}_{2}}-X_{\mathrm{H}}\right)} \beta \gamma.
\end{equation}

The partial derivatives of the mixing entropy are numerically computed.


\end{appendix}


\end{document}